# Optimizing Utility-Scale Solar Siting for Local Economic Benefits and Regional Decarbonization

Papa Yaw Owusu-Obeng [a, *], Steven R. Miller [b], Sarah Banas Mills [c, d], Michael T. Craig [a, e]

[a] School for Environment and Sustainability, University of Michigan, Ann Arbor, MI 48109, USA
[b] Department of Agricultural, Food, and Resource Economics, Michigan State University, Lansing, USA
[c] Graham Sustainability Institute, University of Michigan, Ann Arbor, MI 48104
[d] A. Alfred Taubman College of Architecture and Urban Planning, University of Michigan, Ann Arbor, MI 48109
[e] Department of Industrial and Operations Engineering, University of Michigan, Ann Arbor, MI 48109, USA

*Corresponding author: Papa Yaw Owusu-Obeng
Email: owusup@umich.edu

**Abstract**

The Midwest, with its vast agricultural lands, is rapidly emerging as a key region for utility-scale solar expansion. However, traditional power planning has yet to integrate local economic impact directly into capacity expansion to guide optimal siting decisions. Moreover, existing economic assessments tend to emphasize local benefits while overlooking the opportunity costs of converting productive farmland for utility-scale solar development. This study addresses these gaps by endogenously incorporating local economic metrics into a power system planning model to evaluate how economic impacts influence solar siting, accounting for the cost of lost agricultural output. We analyze all counties within the Great Lakes region—Illinois, Indiana, Michigan, Minnesota, Ohio, and Wisconsin—constructing localized supply and marginal benefit curves that are embedded within a multi-objective optimization framework aimed at minimizing system costs and maximizing community economic benefits. Our findings show that counties with larger economies and lower farmland productivity deliver the highest local economic benefit per megawatt (MW) of installed solar capacity. In Ohio, for example, large counties generate up to $34,500 per MW, driven in part by high property tax revenues, while smaller counties yield 31% less. Accounting for the opportunity cost of displaced agricultural output reduces local benefits by up to 16%, depending on farmland quality. A scenario prioritizing solar investment in counties with higher economic returns increases total economic benefits by $1 billion (or 11%) by 2040, with solar investment shifting away from Michigan and Wisconsin (down by 39%) toward Ohio and Indiana (up by 75%), with only a marginal increase of 0.5% in system-wide costs. These findings underscore the importance of integrating economic considerations into utility-scale solar planning to better align decarbonization goals with regional and local economic development.

**Introduction**

Utility-scale solar photovoltaic (PV) is poised to become the largest contributor to power sector decarbonization, driven by technological advancements, cost reductions, and supportive policies such as the Inflation Reduction Act (IRA). In 2023, the United States added approximately 18.5 GW of utility-scale solar capacity—a 77% increase from the previous year—surpassing wind and other renewable energy technologies in new capacity additions. With over 1,000 GW of new capacity dominating the interconnection queue, utility-scale solar is set to maintain its trajectory of rapid growth in the coming decades. A notable trend in recent years is the surge of proposed projects in the Midwest [1], [2], particularly in states like Ohio and Wisconsin, as land availability and competition shifts deployments away from traditional solar hubs toward emerging energy communities [1]. This regional shift is underpinned by the Midwest's vast rural agricultural lands, characterized by flat terrain and large parcel sizes ideal for hosting utility-scale installations [3], [4], [5]. Notably, between 2012 and 2020, approximately 70% of solar projects in the Midwest were sited on agricultural land [6]. Owing to such land-use advantages the Midwest has emerged as a focal point for future solar development [7].

To support this expansion, generation planning has leveraged optimal utility-scale PV siting that integrate technology potential [8], [9], [10], system cost efficiency [4], [8], [11], [12], [13], regulatory constraints [14], [15] and environmental considerations [16], [17], [18], [19]. More recently, attention has turned to the socio-economic dimensions of utility-scale solar siting, particularly the local economic impacts [5], [20], [21], [22], [23]. These impacts are becoming increasingly important as the growing footprint of solar projects displaces rural agricultural lands [24], [25], creating positive and negative economic spillover effects. On one hand, utility-scale solar can be highly lucrative for local economies. Solar development often provides landowners with significantly higher income through lease agreements compared to traditional agricultural uses. For example, in Michigan, typical solar leases offer landowners an average of $800 per acre annually for terms of 20 to 25 years, significantly exceeding farmland rental rates, which average around $127 per acre per year [26]. Beyond direct landowner benefits, solar installations can generate economic benefits for host communities, including job creation, increased property tax revenues, and long-term economic diversification [5], [20], [22], [23], [27], [28]. On the other hand, the conversion of agricultural lands to solar use can disrupt rural economies by affecting local agricultural supply chains, reducing agricultural employment, and diminishing ancillary industries[5], [29]. Recent studies highlight these nuanced local economic impacts and underscore the importance of evaluating the positive and negative dimensions of solar siting to enhance community acceptance and promote equitable solar deployment [5], [23], [29], [30].

The standard regional economic model for quantifying local economic impacts of new developments is the economic input-output (IO) model. Its development is largely attributed to the Nobel Prize winning economist Wassily Leontief, who developed the concept in the late 1920s and early 1930s [31], while Walter Isard extended its application to localized economic impact analysis [32]. An appropriately-specified IO model makes up the core of the National Renewable Energy Lab's Job and Economic Development Impact Model (JEDI) [33], allowing the user to apply their own region-specific IO model parameters derived from the commercial IO model developed by IMPLAN Group LLC [34] or that provided by the U.S. Department of Commerce's Bureau of Economic Analysis called the Regional Input-Output Modeling System (RIMS II) [35]. IO-based assessments track direct, indirect, and induced economic effects of solar projects, capturing changes in economic indicators such as employment, earnings, gross output, and value-added contributions throughout the project's lifecycle [23]. A growing body of literature documents positive economic outcomes of renewable energy investments at various scales—county [5], [36], [37], [38], [39], state [20], [40], [41], regional [22], [42] and national [28], [43], [44], [45], [46]. Yet, existing studies overlook the economic losses associated with converting farmland to solar. Some researchers argue that to provide a more balanced assessment, models must account for the direct economic effects of lost agricultural production [47]. Ignoring these costs risks overstating net benefits and could hinder efforts to align solar expansion with the long-term sustainability of rural agricultural economies [48].

Another limitation in the current literature is the narrow geographic scope of most economic impact studies. Few U.S.-based analyses of utility-scale solar compare economic outcomes across multiple potential installation sites. Instead, existing studies typically focus on individual projects within a targeted geographic area [21], [27], [49]. This narrow focus, combined with varied assumptions and modeling approaches, makes it difficult to generalize findings or inform broader decision-making. This gap limits our understanding of how economic benefits of utility-scale solar could be leveraged to advance decarbonization goals while enhancing local economic development.

Finally, the current literature also fails to integrate economic impact assessments directly into power system planning or capacity expansion models to guide optimal utility-scale siting. Capacity expansion models typically identify the least-cost mix of generation, storage, and transmission resources necessary to meet electricity demand reliably under various policy constraints. These models have improved spatial resolution to capture resource and technology cost variability, but they generally do not incorporate local economic impacts as endogenous optimization components [8], [14], [17], [50]. Instead, these models rely on a scenario based approach where power system plans are developed first, and only then are separate economic impact evaluations performed [44]. This post-hoc process fails to identify siting configurations that might optimize both technical and economic objectives simultaneously. Given the growing opposition to solar siting, which has led to project delays [51] and increased cost of decarbonization [14], several studies agree that large-scale deployment of utility-scale solar will require community support to ensure successful implementation and operation [52], [53]. Developing siting strategies that benefit rural communities is essential for fostering local acceptance and meeting both solar deployment and broader decarbonization goals.

In this study, we examine local economic effects from utility-solar solar deployment by developing an integrated modeling framework. Our framework is the first to endogenously incorporate local economic metrics into a capacity expansion model. Our analysis focuses on all 524 counties within the Great Lakes region—spanning Illinois, Indiana, Michigan, Minnesota, Ohio, and Wisconsin—and specifically considers photovoltaic (PV) deployment on rural agricultural lands. To quantify county-level economic impacts, we develop a regional economic simulation model employing a set of representative economic IO models. These models simulate how PV solar expenditures circulate through interconnected industries supporting the solar project to generate economic effects that are larger than the direct local expenditures by the PV solar project. The expected economic loss of forgone agricultural production is similarly modeled and subtracted from the positive effects of PV solar expenditures using stylized PV and agricultural expenditure profiles [54]. Therefore, the simulation model provides economy-wide economic impact estimates for each county in the six-state region, where the economic measure is the dollar value of annual transactions generated, the number jobs supported, earnings from those jobs and total regional income (sum of labor and proprietor income and net government revenues from taxes and fees). These estimates cover the full project lifecycle, including installation, operations and maintenance, and decommissioning phases. These net economic gains are represented as marginal benefit curves, detailing the incremental economic benefit per unit of PV capacity added. In parallel, we develop supply curves across counties, informed by earlier work [14], that capture the costs of capacity additions under land-use, zoning, and technoeconomic constraints. We then integrate these curves into a multi-objective capacity expansion (CE) model designed to minimize total annualized system costs while maximizing local economic benefits of utility-scale PV additions. To assess the trade-offs between system costs and economic benefits, we assign varying weights to these objectives and run the CE model from the present through 2040 in five-year increments. Across all scenarios, we implement an 80% $CO_2$ emission reduction target by 2040, aligning with state[55] and federal policies[56] as well as broader decarbonization goals [57].

**Data and Methods**

Our analytical framework is summarized in Fig. 1. First, we conduct site analysis that evaluates county-level land availability for utility-scale PV after applying zoning ordinances and land-use exclusions. We then develop supply curves based on solar resource quality, interconnection costs and land availability, and marginal benefit curves based on county-level economic impacts. These curves serve as inputs to a capacity expansion (CE) model, which optimizes generation and transmission investments.

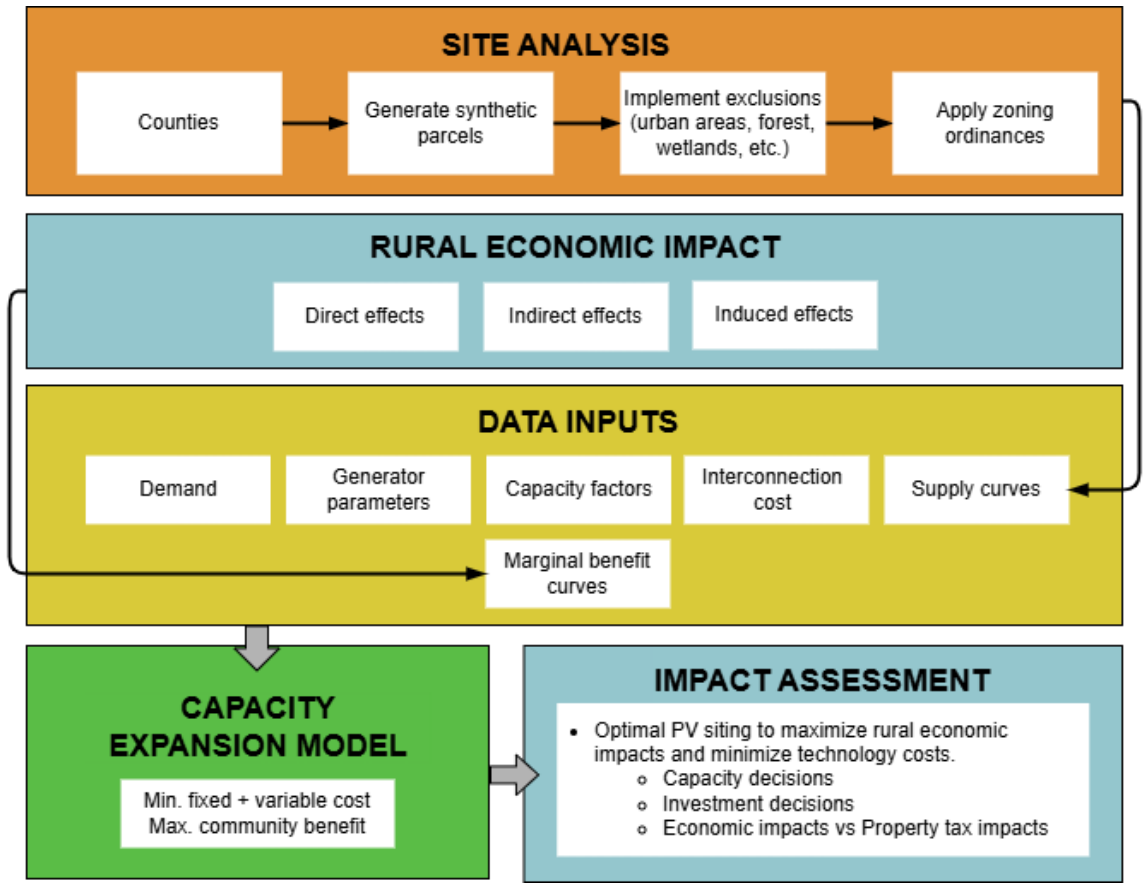

**Fig. 1.** Analytical framework used in this study.

## Quantifying Land Area Available for Utility-Scale PV Development at the County Level

We quantify the land available for utility-scale PV development in each county by applying land-use exclusions and zoning ordinances to individual parcels, following the method outlined in Owusu-Obeng et al [14]. Due to the absence of comprehensive statewide parcel data, we use available parcel data from Wisconsin and Indiana to generate synthetic parcels through geospatial analysis in ArcGIS Pro and nearest neighbor analysis [58], [59] (SI 1, Section 1). We validate our result by comparing the size and distribution of synthetic parcels to the size and distribution of actual parcels the Lawrence Berkeley National Laboratory (LBNL) (SI 1, Fig. S1) [14]. To focus solely on utility-scale PV potential in rural areas and agricultural lands, we exclude urban areas, urban areas buffers, restricted lands, and non-agricultural lands using data on urban land data from the US Census Bureau's, Cropland Data Layer from the United States Department of Agriculture, and Prime Farmland data from the USDA Natural Resources Conservation Service (NRCS) [60], [61], [62] (SI 1, Section 3).

In focusing only on agricultural lands, our approach reflects current trends in utility-scale PV development, which often prioritizes agricultural lands for their flat topography and accessibility [4]. For example, the USDA reports that between 2012 and 2020, the Midwest had the highest proportion of utility-scale solar projects on cropland, with 70% of such projects sited there [6] (SI 1, Fig. S3). While solar development also occurs on non-agricultural lands, we expect a quantitative but not qualitative change in our results, as large-scale deployment on agricultural lands has been and will likely continue to be the dominant form of utility-scale solar deployment in our study region.

We further exclude sites with slope unsuitable for wind or solar development, that is, at or above 10 degrees for solar and 19 degrees for wind [17], [50]. The remaining land parcels are mapped against a database of renewable energy zoning ordinances to evaluate land availability under various zoning regulations [14], [63]. We exclude parcels in zoning jurisdictions that prohibit utility-scale solar PV on agricultural lands. For jurisdictions that permit solar on agricultural lands, we apply road setbacks, participating property line setbacks, non-participating property line setbacks, and minimum and maximum lot size restrictions on individual parcels. We exclude jurisdictions whose zoning ordinances do not mention solar or are "silent" on solar development. In the states included in our study, these "silent" ordinances effectively operate as bans due to a permissive zoning framework, where land uses are prohibited unless explicitly allowed  (SI 1, Section 2). Finally, all available land is aggregated at the county level. For additional modeling details, see SI 1: Site Analysis.

**Assessing Solar Resource at the County Level**

We determine the mean annual hourly capacity factors for each county by integrating meteorological data, such as hourly solar insolation, air temperatures, and other relevant meteorological variables, from the National Solar Radiation Data Base (NSRDB) [64] into the System Advisor Model (SAM) [65]. Furthermore, we estimate spatially explicit costs for connecting utility-scale PV facilities in each county to the bulk transmission grid by measuring the shortest straight-line distance. Our approach accounts for both electrical components and right-of-way costs from the Midcontinent Independent System Operator (MISO) [66], and transmission line data from the Homeland Infrastructure Foundation Level Database [67]. We then combine these interconnection costs with capital and operational cost parameters from the National Renewable Energy Laboratory's (NREL) 2021 Annual Technology Baseline (ATB) [68], using the "Market" case under a Mid cost scenario (SI 1, Table S1), which assumes a moderate pace of technological advancement and cost reduction. Based on these county-level capacity factors and cost estimates, we construct supply curves that serve as inputs to our capacity expansion model.

**Determining the Economic Impact at the County Level**

To estimate county-level economic impacts, we developed a publicly available, spreadsheet-based tool specifically for this study, designed to evaluate utility-scale solar projects across three distinct phases: (1) the short-term installation phase, (2) the long-term operations and maintenance (O&M) phase, and (3) the short-term decommissioning phase at the project's end-of-life. The tool and accompanying documentation are available at [69]. The installation phase is assumed to span 18 to 24 months, while the decommission phase is expected to span less than 12 months. The operating life of the projects is assumed to span 30 years. Key parameters for establishing baselines, such as the project timelines, land requirements, and cost structures are derived from NREL's ATB [68] and from survey-based economic impact estimates collected by Lawrence Berkeley National Laboratory [70]. For additional modeling details, see SI 2: PV Solar Net Economic Impact Estimation Model.

For each project phase, direct project expenditures drive economic impact estimates. For the installation phase, total installation expenditures are distributed to respective commodities based on ATB aggregates (e.g., land leases, site preparation, equipment procurement, installation, and grid interconnection). Because not all expenditures are captured within the county of installation, local purchase percentages (LPPs) are applied to each commodity to estimate the share of spending retained locally. For example, expenditures on solar panels are assumed to have no local capture, whereas most construction-related expenditures are retained within the local economy. Likewise, while electrical components and fixtures are sourced externally, services like ground maintenance (e.g., mowing) are assumed to be entirely local. Administrative and monitoring expenditures are assumed to occur outside the county, while ongoing facility services are mostly captured locally. O&M expenditures are assumed to increase as equipment ages. To account for this, annual costs for depreciable O&M categories (e.g., equipment maintenance and repair) are modeled with a 5% annual escalation rate. Decommissioning costs, covering the removal of infrastructure and land restoration, are estimated on a per-MW basis. All expenditures are measured in constant 2024 dollars to avoid inflation projections.

Land lease payments occur through all three project phases, based on project acreage at a rate of $580 per acre, measured in 2024 dollars. These payments drive household spending and are modeled as induced effects using household expenditure distributions from the U.S. Department of Labor Consumer Expenditure Survey [71] and RIMS II multipliers [35]. We also model annual property tax liabilities based on state-specific PV taxation policies. While each county may have different rates, the basis of taxation is generally the same for all counties within a state. Average county millage, or sales/excise tax rates were applied, respectively. All property tax revenues were modeled as government spending, and also as induced effects, following applicable depreciation rules for property taxation in each state.

Across all phases, agricultural production will be disrupted on the share of project acres expected to be converted to solar. Heuristically, that is set to 80 percent of total project acres, where the total project acres affected depends on the project size. We assume that the displaced agricultural acres would otherwise grow a rotation of corn, soybean or wheat, reflecting common three year-crop rotations in the Great Lakes Region. Recent Ohio State University Extension crop enterprise budgets [72] were used to develop a profile of per acre revenues and expenditures as baselines agricultural output that will be supplanted by the PV project, under each of the three crops. We assume that one-third of the impacted acreage would be assigned to each crop annually, allowing us to calculate a composite average annual loss in agricultural output. These values are assumed to be constant over time and measured in constant 2024-dollar values. To account for spatial differences in productivity, we classify counties into high-, medium-, or low-yield categories using 2017 USDA National Agricultural Statistics Services (NASS) county-level yield data for the three crops [54]. If at least two crop yields in a county exceed the six-state average, the county is deemed high-yield; if at least two fall below, it is low-yield; otherwise, it is average. Agricultural losses are adjusted accordingly: +10% for high-yield counties and –10% for low-yield ones.

We estimate total economic impacts using RIMS II input-output (IO) multipliers, which capture direct, indirect, and induced effects for each project phase. Multipliers vary based on local supply chain strength and county economic size, with larger or more diverse economies capturing more internal transactions. Counties are grouped into three population-based size categories—small (≤15,000), medium (≤30,000), and large (≤50,000)—excluding metropolitan areas (S2, Figure S1) [73]. One representative county per size and state (6 states × 3 sizes) results in 18 county profiles. RIMS II multipliers were acquired for each and mapped to 372 industry categories [74]. All positive economic effects (from construction, O&M, decommissioning) are estimated using these multipliers. Negative effects from displaced agricultural output are modeled similarly by applying Type II multipliers to the lost direct agricultural revenue. Together, these estimates yield a net economic impact for each county profile over the full project lifecycle.

**Capacity Expansion Model**

We employ a long-term capacity expansion (CE) model to identify optimal utility-scale PV investment in each county. The CE model simultaneously optimizes investments in generation and transmission infrastructure, as well as hourly system operations. The model's objective is to minimize total annualized system costs while also maximizing net county-level economic impacts from utility-scale solar deployment. To balance these dual objectives, we apply a weighting factor ranging from 0 to 1, capturing the trade-offs between cost minimization and economic benefit. A higher weight on the cost objective results in the model prioritizing system cost reductions, while a lower weight places greater emphasis on maximizing local economic impacts. Intermediate weights balance both objectives equally.

Total system costs comprise the sum of annualized fixed investment and variable operating cost. Fixed costs account for capital and fixed operations and maintenance (O&M) costs of new transmission lines, electricity generators, and energy storage systems. Variable costs account for fuel and variable O&M costs incurred by both new and existing units. To capture local economic benefits, we incorporate county-level metrics for the total value-added contributions from solar development (see Data and Methods: Determining the Economic Impact at the County Level and S1 2).

Following NREL's Regional Energy Deployment System (ReEDS) Model, our study region is subdivided into 24 subregions, between which limited transmission capacity exists [8]. At each time step, the model selects new investments from a portfolio that includes county-level solar PV and wind, as well as subregional investments in coal steam plants with carbon capture and sequestration (CCS), natural gas combined cycle (NGCC) facilities with and without CCS, nuclear power stations, battery storage systems, and transmission infrastructure. For solar capacity, we determine an upper bound in each county based on the extent of agricultural land available after the application of land exclusions and zoning ordinances. Similarly, wind is confined to agricultural lands. We convert land area to potential capacity for wind and solar using wind and solar power densities of 0.5 $Wm^{-2}$ and 5.4 $Wm^{-2}$, respectively [75]. We obtain future capital and operational costs from NREL's Annual Technology Baseline, consistent with inputs used in our economic model (see prior section)[68]; interregional transmission capacity from ReEDS [8]; and future electricity demand profiles by subregion for a moderate electrification future [76]. We account for the Inflation Reduction Act by applying a 30% Investment Tax Credit (ITC) to the capital costs of solar and wind. We assume that all eligibility requirements for the 30% ITC, such as prevailing wage, project size and apprenticeship provisions, are fully met. We do not incorporate additional tax credit adders (e.g., for energy communities) to ensure a conservative representation of solar and wind capital costs across the study region. Wind, solar, and demand timeseries correspond to 2012 meteorology, capturing co-variability in these meteorologically-dependent variables. A list of our input datasets and assumptions is detailed in SI 1, Table S1. Our initial generation fleet is detailed in S1, Table S2.

The CE model enforces several system-level constraints, including balancing regional electricity and demand and meeting reserve requirements on an hourly basis (SI 3, Section 2). The model imposes unit-level constraints, such as upper limits on individual power plant capacities (SI 3, Section 3). Power flows between load regions are represented using a simple transport formulation, which determines optimal interregional energy transfers subject to maximum transmission capacities [12]. For computational tractability, each modeled year is condensed into two representative days per season plus the day of peak annual demand [77] (SI). The CE model is implemented in GAMS and solved with the CPLEX solver. For additional modeling information see SI 3 and SI 4.

**Scenario Framework**

To explore the tradeoffs between local economic benefits and system cost minimization in utility-scale solar deployment, we design a series of scenarios that adjust the relative weighting of our two objectives (minimizing total system costs and maximizing net county economic impacts). Specifically, we vary the importance placed on local economic outcomes and total system costs by applying a set of weights that range from 0 to 1 in increments of 0.2. At one extreme, the model prioritizes only economic benefits (100% weight on local benefits and 0% on cost), while at the other, it focuses on minimizing costs (0% weight on local benefits and 100% on cost). Intermediate scenarios allocate partial weights to each objective to balance local economic development with cost-effectiveness. All scenarios are evaluated under a consistent set of policy and decarbonization targets. We impose an 80% reduction in $CO_2$ emissions by 2040, aligned with state[55] and federal [56] policies, as well as broader decarbonization goals[57] to achieve net-zero emissions by 2050 [57]. The scenarios are run from present through 2040 in five-year increments, allowing us to track the evolution of generation portfolios, transmission infrastructure, and local economic outcomes over time.

**Results**

**Drivers of local economic impact and effect of lost agricultural output**

To quantify the extent, distribution, and drivers of local economic impacts, we first examine the annual net economic benefits from the value-added contributions per megawatt (MW) of installed solar capacity at the county level (Fig. 1),  alongside the economic losses from farmland conversion. Figures 1c reveal substantial variation in value-added across and within states, driven primarily by the size of the local economy (Fig. 1a), and to a lesser extent by the productivity of the agricultural land converted to solar (Fig. 1b, SI-5 Fig. 1).

Regionally, the size of the local economy emerges as the dominant factor shaping net impacts, with counties in large, medium, and small economies experiencing average annual benefits ranging from $23,400–$34,600, $20,900–$29,700, and $18,500–$23,000 per MW, respectively. Larger economies tend to include a wider array of endogenous industries, enabling greater retention of economic activity and thus higher value-added. In contrast, smaller economies are more dependent on external supply chains, reducing local capture of expenditures. This regional pattern is also evident at the state level (Fig. 2). For example, Ohio shows the highest economic impacts, with average county-level values of $34,500, $29,700, and $20,000 per MW-year for large, medium, and small counties, respectively. By contrast, Indiana represents the lower bound, with corresponding values 31%, 18%, and 6% lower than those in Ohio.

Accounting for the loss in agricultural productivity from farmland conversion reduces economic benefits. Regionally, net economic impacts decline to $19,000-$30,600, $17,300-$25,700, and $14,500-$19,400 per MW-year for large, medium, and small counties, respectively. In Ohio, average reductions in value-added amount to 12%, 15%, and 16%—or $4,400, $4,500, and $3,200 per MW-year—for large, medium, and small economies, respectively (Fig. 2). The magnitude of reductions in large and medium counties is similar due to the predominance of land types with comparable average farmland productivity. When evaluating agricultural productivity losses independently of economic size, counties with "above average" productivity experience the highest reductions (averaging $4,700 per MW-year), compared to $4,200 and $3,500 for counties with average and below-average productivity, respectively.

The impact of net property tax revenue to local governments—reflecting the replacement of lost agricultural taxes with gained solar-related property taxes—varies by state (SI-5, Table S1). Ohio leads, with the highest net present value (NPV) of property tax revenues at $103,400 per MW, driven by its Payment In Lieu of Taxes (PILOT) program, which generates approximately $8,750 per MW annually and contributes to the state's overall highest economic impact (Fig. 1c). Wisconsin follows with an average NPV of $60,300 per MW, while Indiana and Michigan report more modest values of $39,400 and $38,800 per MW, respectively. Illinois yields $28,000 per MW, and Minnesota ranks lowest, with a property tax NPV of just $19,700 per MW.

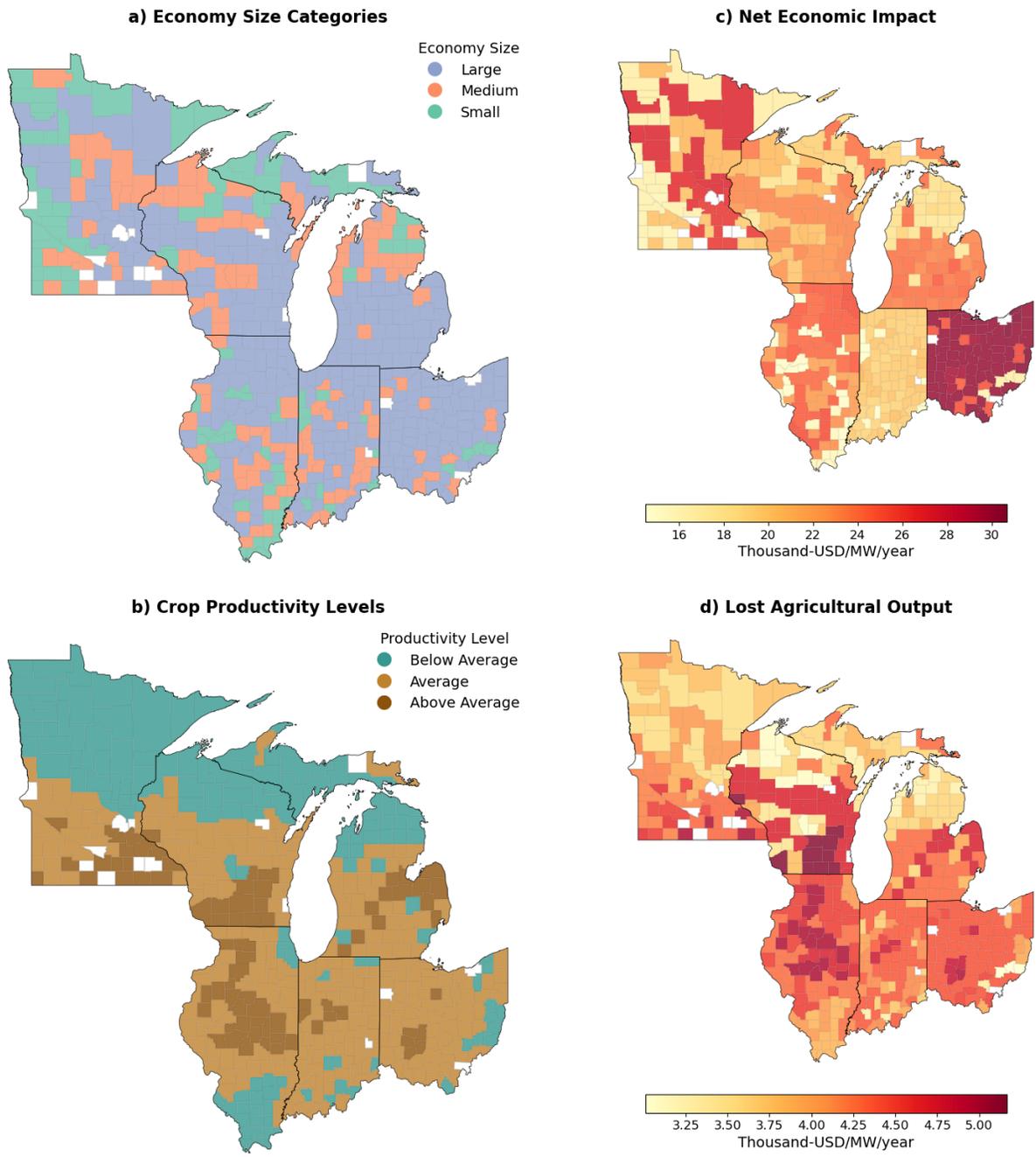

Fig. 1. County-level classification of (a) economy size (large, medium, small) and (b) crop productivity (below average, average, above average), along with (c) net economic impact per MW of installed solar capacity, accounting for the loss of agricultural output, and (d) annual economic losses from displaced farmland. Darker shades in (c) and (d) indicate higher value-added contributions and greater agricultural losses, respectively.

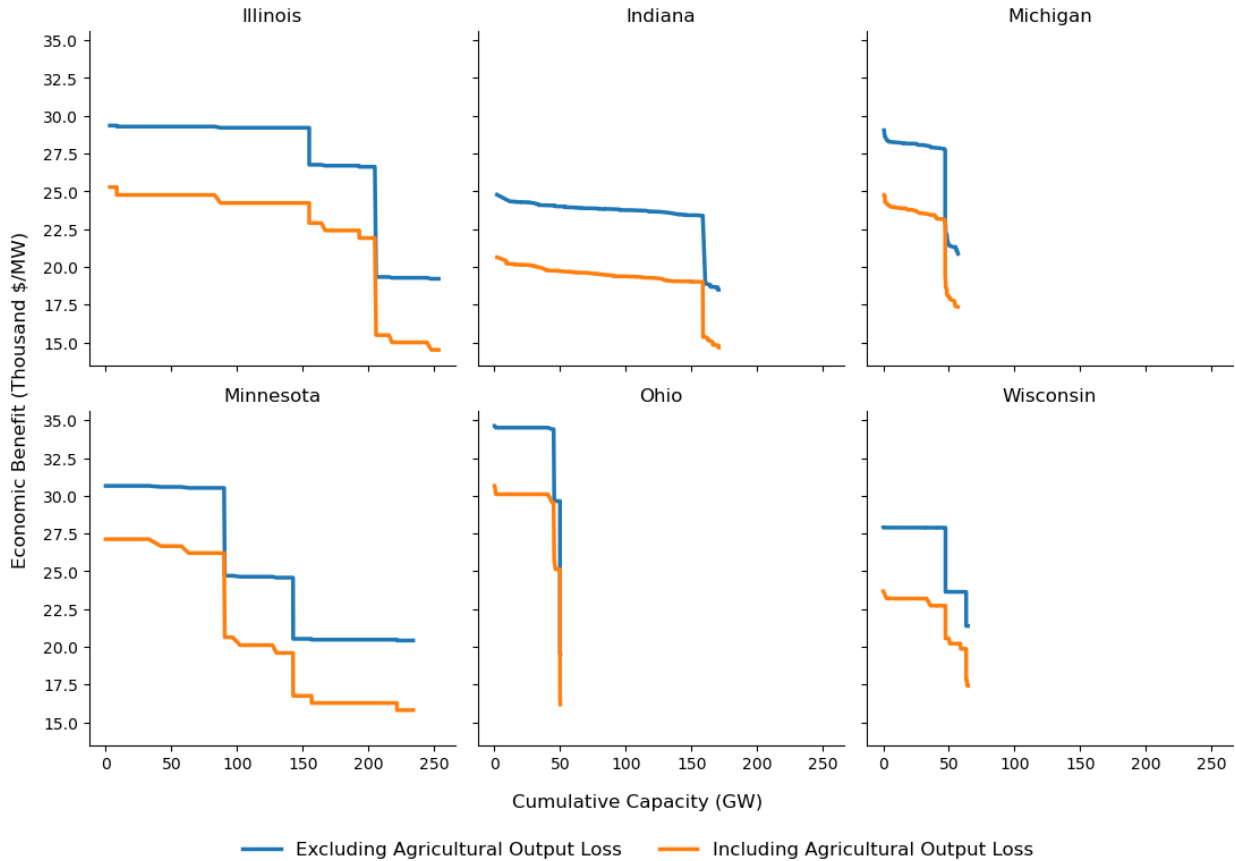

Fig. 2. Marginal benefit curves showing the value-added (thousand $/MW) as a function of cumulative solar capacity (GW). The blue line represents value-added without accounting for agricultural productivity losses, while the orange line accounts for agricultural productivity losses. The divergence between the blue and orange curves highlights the impact of agricultural productivity losses on economic benefits.

**Effect of local economic impacts on utility-scale PV capacity investments**

Fig. 3. illustrates investment decisions from our CE model by 2040 across the study region (3a) and their distribution by state (3b) under the base case scenario, which assumes a 100% weighting for minimizing system costs. Under this scenario, solar accounts for 37% (64 GW) of all new investments (Fig. 3a). These investments are driven primarily by zoning regulations, resource availability, and state-level minimum generation requirements. For instance, Ohio sees significant solar deployment in its eastern regions, despite lower solar capacity factors, largely due to the absence of restrictive zoning ordinances in these areas (SI-1 Fig. S4).

In Fig. 3c, we hold total investment capacity constant at the level observed in the base (100% cost) scenario and vary the weighting between cost minimization and community benefit maximization. This allows us to examine how capacity reallocates across counties with differing economy sizes and farmland productivity levels. Without imposing an upper limit on total buildout, the benefit-maximizing objective leads to unrealistically large solar investments (an infinite amount of investment, or an unbounded optimization problem, at the limit of only including maximizing community benefits in the objective via a weight of 1). Prioritizing community benefits shifts solar deployment to counties with larger economies and below-average farmland productivity. For instance, counties classified as large economies receive a total solar investment of 36, 41, and 52 GW when community benefit weight is 0%, 50%, and 80%, respectively. This increase in investment comes at the expense of investments in medium and small counties declining. Larger economies attract more investment because they yield higher local value-added impacts, due to their broader and more diverse industrial bases. From a system perspective, these locations may be less cost-effective in terms of resource quality, but the associated economic benefits more than offset the higher technology deployment costs under a community benefit–oriented objective. This shift is also influenced the availability of land, which is restricted by zoning regulations.

The influence of county farmland productivity plays a smaller role than county economy size. For example, solar investment in below-average productivity counties increases by 11%, from 16 GW under the 0% community benefit scenario to 19 GW under the 80% community benefit scenario, while investments in average and above-average productivity categories decline by 14% and 17%, respectively. This trend reflects the model's preference to avoid converting high-value farmlands when local economic benefits are prioritized.

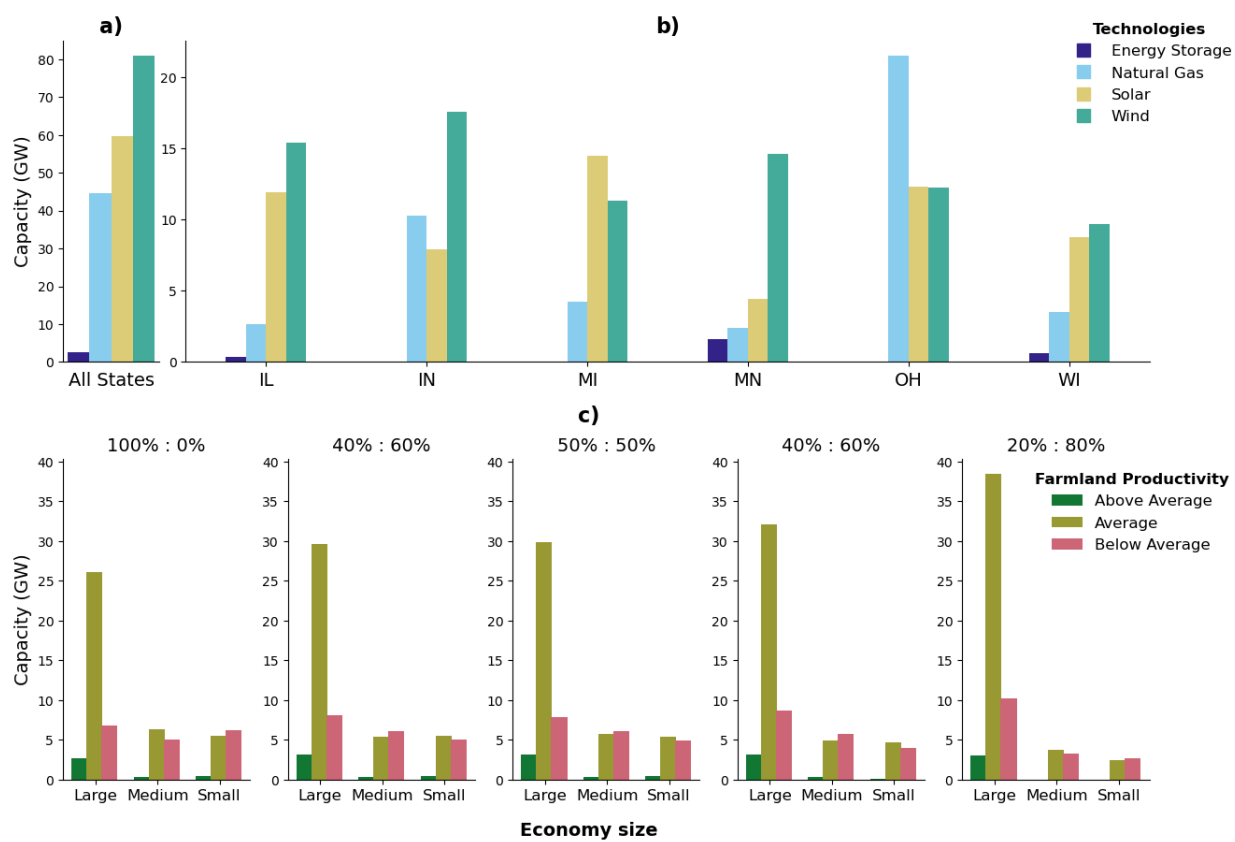

Fig. 3: Allocation of new technology investments by 2040 across the study area and states, and their distribution by economic and farmland productivity categories. (a) Total capacity additions by technology type across all states under the cost-minimization-only scenario (100% weighting of costs). (b) Distribution of investments across individual states under the same scenario. (c) Reallocation of solar capacity investments across county economy size and soil productivity categories for varying weighting ratio of cost minimization to community benefit maximization (written as cost minimization weight : community benefit weight).

**Effect of local economic impacts on geographical distribution of new utility-scale solar**

Fig. 4 illustrates the geographic distribution of new utility-scale solar deployments as community benefit weighting increases. Capacity shifts within and across states are influenced by economy size and farmland productivity categories. Comparing the base (100% cost) scenario to the 80% benefit scenario, new solar investments shift from Michigan (down 5.8 GW or 39%) and Wisconsin (down 2.7 GW or 31%) to states with a higher prevalence of counties characterized by large economies and lower soil productivity, increasing investments in Ohio (up 9 GW or 75%), Illinois (up 2.4 GW or 20%), Indiana (up 780 MW or 10%), and Minnesota (up 380 MW or 6%) (Fig. 4). These shifts also lead to the emergence of new counties with solar investments, including 9 additional counties in Ohio (18% increase), 8 in Illinois (35% increase), 2 in Indiana (14% increase), and 1 in Minnesota (8% increase).

Accompanying these shifts is a complementary increase in wind capacity and a decline in natural gas capacity. In the 80% benefit scenario, wind capacity grows significantly in Minnesota (by 6.9 GW or 45%) and Wisconsin (by 2.3 GW or 23%) (refer to Fig. 4 color bar). This growth is attributed to changes in solar generation profiles at new sites, requiring additional investment to ensure continued balance between electricity demand and supply. Conversely, total natural gas capacity decreases marginally by 2.1 GW (5%), with 2 GW occurring in Ohio.

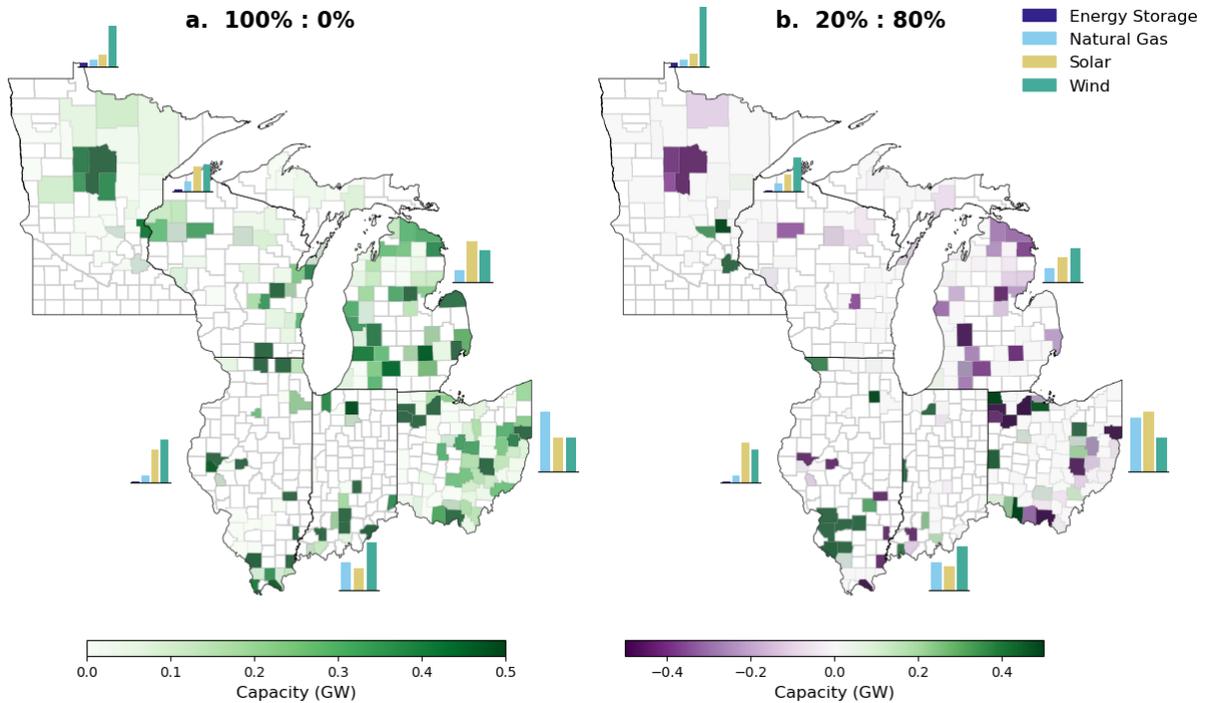

Fig. 4. Geographic shifts in new utility-scale solar as the weighting ratio between cost to local economic benefit is adjusted. (a) Is the represent the 100% cost scenario and (b) represents the 100% benefit scenario. Positive values on the scale (green) represent capacity additions, while negative values (purple) indicate capacity reductions. Color bars indicate the total capacity investment for each technology, with yellow for solar, green for wind, purple for energy storage, and blue for natural gas.

**Effect of local economic impacts on the costs of utility-scale solar investments.**

Prioritizing community benefits in siting decisions significantly increases local economic impacts with minimal effect on system-wide costs. As shown in Table S2 (SI-5), shifting from 0% to 80% community benefit weighting raises total value-added from $8.7 billion to $9.7 billion—a 11% or $1 billion gain— while system-wide investment rises by less than 0.5%. The magnitude of these local economic benefits varies widely across states. Minnesota experiences the largest percentage increase (41%, $442 million), followed by Ohio (24%, or $520 million), Indiana (5%, $75 million), Illinois (3%, $44 million), and Wisconsin (1%, $16 million). In contrast, Michigan experiences a reduction of 7% ($100 million). These reductions reflect their lower value-added contributions per unit of installed capacity (Fig. 1d).

**Discussion**

This study quantifies the effect of local economic impacts on utility-solar investment and power sector decarbonization in the Great Lakes region of Illinois, Indiana, Michigan, Minnesota, Ohio, and Wisconsin. We develop and integrate marginal benefit curves of installing solar alongside supply curves for each county into a power system planning (or capacity expansion) model to simultaneously minimize system costs and maximize local economic benefits. We explore various weighting scenarios to assess the trade-offs between cost minimization and benefit maximization.

Our analysis reveals that relocating solar investments towards counties offering higher net economic benefits does not increase the total cost of decarbonization. Across our region, prioritizing local solar economic benefits increases aggregate economic impact by up to 11% ($1 billion) while total decarbonization costs increases by less than 0.5%. This finding challenges the conventional assumption that maximizing local economic benefits from renewable energy projects necessitates significantly higher overall system costs. This result aligns with emerging research on spatially explicit renewable energy planning [78], [79], demonstrating that strategic siting can unlock co-benefits without compromising cost-effectiveness. By adopting targeted planning guidelines, developers can prioritize counties with higher net economic benefits, fostering community acceptance and reducing local opposition or delays associated with permitting.

We also find that counties with robust local economies and lower farmland productivity deliver the highest value-added per megawatt (MW) of installed capacity. For example, large economies in Ohio generate an average of $34,500 per MW-year, a value that is 42% higher than that observed in smaller economies within the same state. Accounting for the opportunity cost of agricultural losses reduces these value-added contributions by up to 16% in some counties, depending on farmland productivity. This correlation between economic size and impact, while varying in magnitude across counties and states due to differing economic multipliers, remains consistent. This is expected, as previous studies using input-output (I-O) economic multipliers [5], [36], [39], also demonstrate a proportional scaling of economic impact with economy size. However, these prior studies often present an overoptimistic view by capturing solely positive economic benefits. Our findings highlight the critical importance of incorporating the opportunity cost of agricultural land into local economic assessments of solar development. Local governments should recognize that these costs can offset the perceived benefits, particularly in counties where agriculture contributes significantly to local revenues. While policies like agrivoltaics may reduce these opportunity costs, the required investment in specialized racking and mounting increases system costs relative to traditional utility-scale solar [80]. By explicitly weighing

foregone agricultural revenues against projected economic gains, policymakers and community stakeholders can better target solar investments to regions where total net benefits are maximized.

Further, prioritizing locations with higher economic impacts leads to shifts in utility-scale solar capacity within and across states, resulting in a net reduction in natural gas capacity and a corresponding increase in wind capacity. Specifically, shifting 8.5 GW of solar investment from Wisconsin and Michigan to the remaining states increases wind capacity in the recipient states and concurrently reduces natural gas generation across the system. This technology substitution highlights the interconnected nature of renewable energy resource planning and the importance of system-wide coordination when evaluating local economic benefits. A regionally coordinated approach to solar deployment, where states collaborate to identify optimal siting locations based on both local economic and system-wide criteria, could diversify renewable energy resources, enhance grid reliability, and reduce reliance on fossil fuels.

Finally, we find that net property tax payments and payments in lieu of taxes (PILOT) are among the most influential drivers of local economic impact that are directly controllable by local governments. Our results show that Ohio's PILOT program, which provides relatively higher payments, correlates with a 75% increase in solar investment under scenarios that prioritize local economic benefits. While such payments can generate substantial revenue for local jurisdictions [81], excessively high or poorly structured tax rates may deter solar development. Previous studies have highlighted this trade-off, noting that elevated local tax burdens can raise project costs, reduce the net present value for investors, and shift development to neighboring areas with more favorable tax policies [82]. Therefore, local governments must strike a balance between maximizing public revenue and maintaining tax structures that remain attractive to private investment.

Our study has several limitations that future research could examine. First, due to the complexity and variability of county-level tax structures, we compiled property tax data at the state level rather than the county level. This aggregation may mask important local variations and potentially lead to under- or overestimation of property taxes in specific counties. Future research incorporating more granular, county-specific property tax data would provide a more precise assessment of local economic impacts. Second, while grouping counties under broad categories of economy size and agricultural productivity allowed us to assess the influence of these factors on local economic impacts and enabled generalization of our results, this categorization could obscure finer-scale relationships that might emerge from a more disaggregated analysis. Third, although our model integrates zoning and economic impact data for solar development, we lack comparable data for wind and other technologies. As a result, scenarios that optimize solely for cost may underestimate the siting challenges associated with wind or storage, while scenarios that prioritize local economic benefits may inadvertently favor more solar buildouts. To mitigate this bias, we hold solar capacity constant in scenarios that prioritize economic impacts, fixing it at the levels identified under cost-only optimization. Nonetheless, future work should seek to develop comparable zoning and local economic impact data for all technologies to enable a more balanced and comprehensive planning framework.

**Acknowledgments**

This material is based upon work supported by the U.S. Department of Energy's Office of Energy Efficiency and Renewable Energy (EERE) under the Solar Energy Technologies Office Award Number DE-EE0009361. The views expressed herein do not necessarily represent the views of the U.S. Department of Energy or the United States Government.

Optimizing Utility-Scale Solar Siting for Local Economic Benefits and
Regional Decarbonization:

Site Analysis

Papa Yaw Owusu-Obeng [a, *], Steven R. Miller [b], Sarah Banas Mills [c, d], Michael T. Craig [a, e]

[a] School for Environment and Sustainability, University of Michigan, Ann Arbor, MI 48109, USA
[b] Department of Agricultural, Food, and Resource Economics, Michigan State University, Lansing, USA
[c] Graham Sustainability Institute, University of Michigan, Ann Arbor, MI 48104
[d] A. Alfred Taubman College of Architecture and Urban Planning, University of Michigan, Ann Arbor, MI 48109
[e] Department of Industrial and Operations Engineering, University of Michigan, Ann Arbor, MI 48109, USA

**This PDF file includes:**

> Supporting text
> Figures S1 to S4
> Table S1
> SI References

**Supporting Information Text**

**1. Generating synthetic parcels**

Zoning ordinances operate at the parcel level, so accounting for zoning ordinances requires individual parcels' boundaries, size, and geographical location. Comprehensive statewide parcel data for four of our six study states (Illinois, Michigan, Minnesota, Ohio) is not publicly available. To resolve this, we generate synthetic parcels for each county subdivision by utilizing parcel shapefiles for Wisconsin [1] and Indiana [2], along with county subdivision shapefiles in our study region. Subdivision shapefiles are obtained from the US Census Cartographic Boundary Data [3]. Using ArcGIS [4], we merge subdivision and parcel polygons in Wisconsin and Indiana. This enables us to assign parcels to their respective subdivisions. The analysis yields approximately 7.14 million parcels within a total of 2,547 subdivisions in Wisconsin and Indiana. We estimate the length, width, and area dimensions of individual subdivision polygons. Having the boundaries of county subdivisions and their containing parcels, we utilize this data as our sample to generate parcels for the remaining 7966 unique county subdivisions in our study region with no parcel information.

To generate synthetic parcels, we utilize the nearest neighbor analysis to map each population (Illinois, Michigan, Minnesota, Ohio) subdivision to the closest sample (Wisconsin and Indiana) subdivision based on their shortest Euclidean distance between their length, width, and area dimensions [5]. We establish a threshold on the Euclidean distance to ensure that larger county subdivisions in our population were iteratively matched to the closest sample until the remaining land area met the threshold. Once each population was matched to a set of sample county subdivisions, we mapped the parcels in the sample to the matched population to generate synthetic parcels for each county subdivision. To validate our result, we compare the distribution of synthetic parcel sizes with the distribution of actual parcel sizes. Despite not having access to actual parcel data, we acquire 10 equal percentile distributions of actual county subdivision parcel sizes from the Lawrence Berkeley National Laboratory (LBNL) for our analysis. We ranked our synthetic parcels into 10 equal percentiles and juxtaposed our distribution with those from LBNL. Fig. S1 presents a comparison of the synthetic and actual parcel distributions for specific states in our study region. Our analysis demonstrates that our method accurately captures the size and distribution of actual parcels.

To focus solely on utility-scale PV potential in rural areas, we exclude urban areas along with a 0.5 km buffer within each subdivision using shapefiles of urban lands from the US Census Bureau's TIGER/Line Data Files [6].

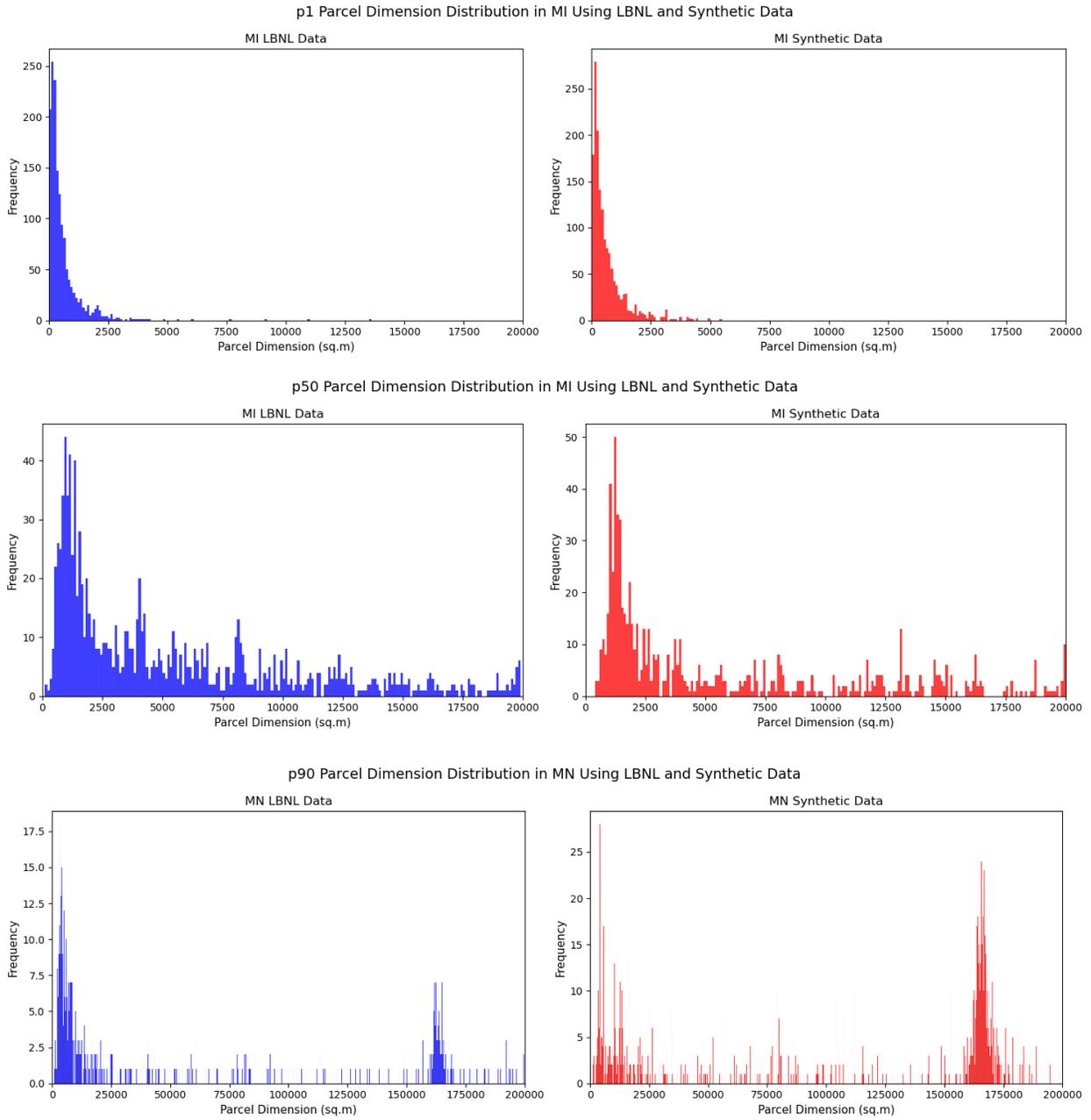

**Fig. S1.** presents a side-by-side comparison of parcel distributions between synthetic and actual data for Michigan and Minnesota. The percentiles P1, P50, and P90 represent the areas of parcels at the 10th, 50th, and 90th percentiles, respectively.

## 2. Implementation of zoning ordinances on synthetic parcels

We map subdivisions enclosing our parcels to a database of renewable energy zoning ordinances to identify jurisdictions that regulate principal-use solar energy systems (SES). Principal-use, within the scope of our study, refers to the generation and export of electricity from utility-scale PV facilities. Zoning regulations can either allow, prohibit or omit the deployment of principal-use solar energy systems within a subdivision. We obtain zoning ordinances from the University of Michigan's Graham Sustainability Institute [7], which host an open-access database comprising over 2,474 local zoning jurisdictions in our study region. These zoning ordinances are publicly accessible through an online database [7]. For jurisdictions allowing principal-use solar energy systems, the database records specifications for size, density, and setback requirements that dictate system layout [7].

We note that in the 6 states in our study region, there are over 2,474 local zoning jurisdictions. In some states (IL and IN), zoning in rural areas is conducted entirely at the county level. However, in Minnesota, Wisconsin, Michigan, and Ohio, there is a combination of township- and county-zoned jurisdictions, as well as townships that are unzoned (unincorporated areas) --where there are no land use regulations for solar, or any other land uses.  Commonly, county governments across the country are responsible for zoning in the unincorporated areas of the county. Thus, in cases where the county holds zoning authority, we apply the county zoning laws to their respective unincorporated areas. By doing so, we utilize the 2,474 local zoning jurisdictions in our database to characterize the status of 9,997 county subdivisions.

In our analysis of 9,997 zoned county subdivisions (i.e., subdivisions have zoning regulations), we found 4,417 subdivisions to have established regulations on principal-use solar energy systems while 3,262 lack such regulations. Among subdivisions that regulate principal-use, 512 subdivisions have outright bans on large scale solar deployment in their dominant agricultural district, while 3,905 allow deployment in agricultural districts. Fig. S2 illustrates the distribution of county subdivisions that either allow or ban solar.

For jurisdictions allowing solar, we apply subdivision-specific regulations to each parcel within that subdivision to quantify each parcel's developable land area. We then sum developable land by parcel back to the subdivision level. While utility-scale photovoltaic (PV) systems typically occupy multiple parcels, we refrain from merging parcels before applying regulations, as regulations including setbacks are enforced on individual parcels irrespective of the number of participating lots.

We first conduct spatial analysis in ArcGIS pro to determine the average distribution of parcel boundaries adjoining different land types for physical solar facilities, specifically roads (19%), participating properties (44%), and non-participating properties (37%). The same distribution is used to segment our developable land parcel boundaries into corresponding land-use classifications. For segments bounded by the road, we implement the road setback regulation by subtracting the specified setback distance from the parcel's edge towards its center. We adopt a similar approach to implement setback regulations related to participating properties (PPL) and non-participating properties (NPPL) for segments bounded by these respective land types.

Finally, we enforce a minimum (MinLS) and maximum (MaxLS) lot size requirement to exclude parcels falling outside the specified limits for large-scale solar energy systems. After implementing regulations, our results yield 3821 out of 3905 subdivision that allow solar in their agricultural district. This represents a 30% reduction in land area.

For 1,340 unzoned subdivisions (i.e., subdivisions that do not have zoning regulations) that allow unrestricted utility-scale PV development, we quantify the available land area in these subdivisions by enforcing a set of standard land-use regulations employed agricultural districts. We implement a 100ft road setback, 50ft property line setback, a 40% maximum lot size coverage and 1-acre minimum lot size requirement. This process yields an additional 1,340 subdivisions allowing solar as shown in Fig. S2.

Our database captures a snapshot in time of utility-scale PV zoning ordinances. Utility-scale PV projects can incentivize the revision of zoning ordinances to either allow or restrict solar energy in a jurisdiction [8]. In this context, we predict the evolution of 3262 zoned subdivisions currently lacking solar provisions in their ordinances by extrapolating the distribution of solar-allowing and -restricting subdivisions from the 4,417 subdivisions already having provisions for principal-use of solar energy. This distribution is represented in Fig. S2. Similarly, we enforce the setback distances and lot size requirements to parcels using representative values.

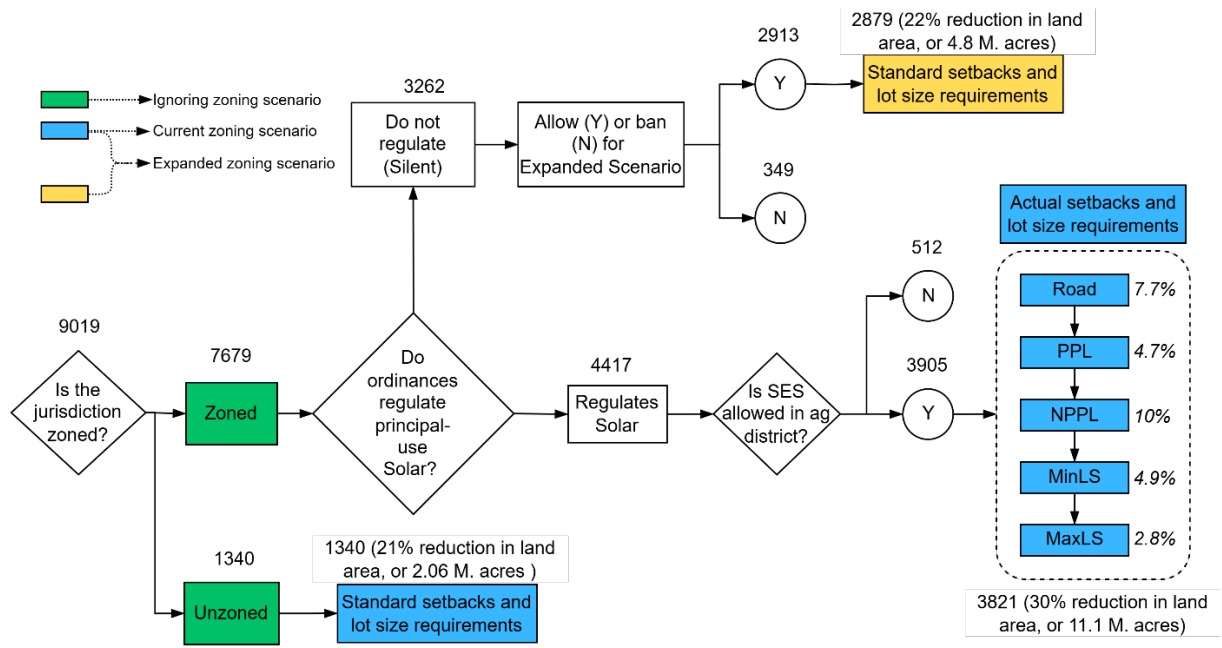

**Fig. S2.** Mapping of zoning regulations for principal-use solar energy systems, or utility-scale PV systems. Subdivisions that allow solar on agricultural lands are labeled as "Y" and those that prohibit solar on agricultural lands are labeled as "N". Setbacks from roads are labelled as "Road", those from participating property lines as "PPL", and from non-participating property lines as "NPPL". Minimum and maximum lot size requirements are labelled as "MinLS" and "MaxLS", respectively. The count tracks the number of subdivisions where specific regulations are applied, while the percentages indicate the land area reduction from regulations that limit lot size. Boxes

are shaded by the scenario in which subdivisions are included, as detailed in the Scenario Framework section (green for ignoring zoning, blue for current zoning, and a combination of yellow and blue for expanded zoning).

### 3. Quantifying available agricultural land area for PV deployment

Using ArcGIS and cropland and land cover data layers [9], [10], we assign each 30-meter pixel in each subdivision to one of three distinct categories: agricultural lands which encompasses prime farmlands used for crop production [9]; non-agricultural lands, which include barren, shrub and grasslands; and restricted lands, which include areas that are unsuitable for utility-scale PV development, such as wetlands, water, ice, forests and impervious surfaces. Our findings indicate that subdivisions in our study region predominantly consist of agricultural lands compared to non-agricultural lands [11]. To gain insight into the placement of utility-scale PV facilities, we utilize data from EIA-Form 860 [12] and geolocate existing PV facilities in our study region. By creating buffers around each facility and analyzing the pixel density and characteristics, we find that 86% of utility-scale facilities within our study region are located predominantly on agricultural lands, with the remaining on non-agricultural lands, while deployment is totally avoided on restricted lands. Based on this observation, we apply the proportion of agricultural landcover in each subdivision to the total land available after implementing zoning ordinances. Our results yield the developable agricultural land areas in each subdivision for utility-scale PV deployment.

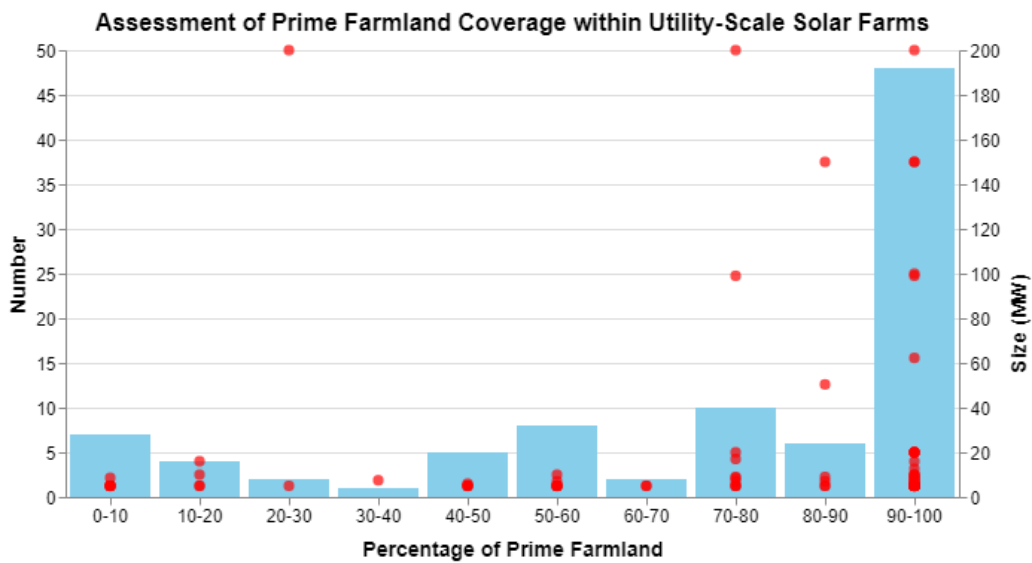

**Fig. S3.** Assessment of prime farmland coverage on utility-scale solar facilities. The blue bars represent the number of utility-scale solar facilities within different percentage ranges of prime farmland coverage, while the red dots correspond to the size of the facilities.

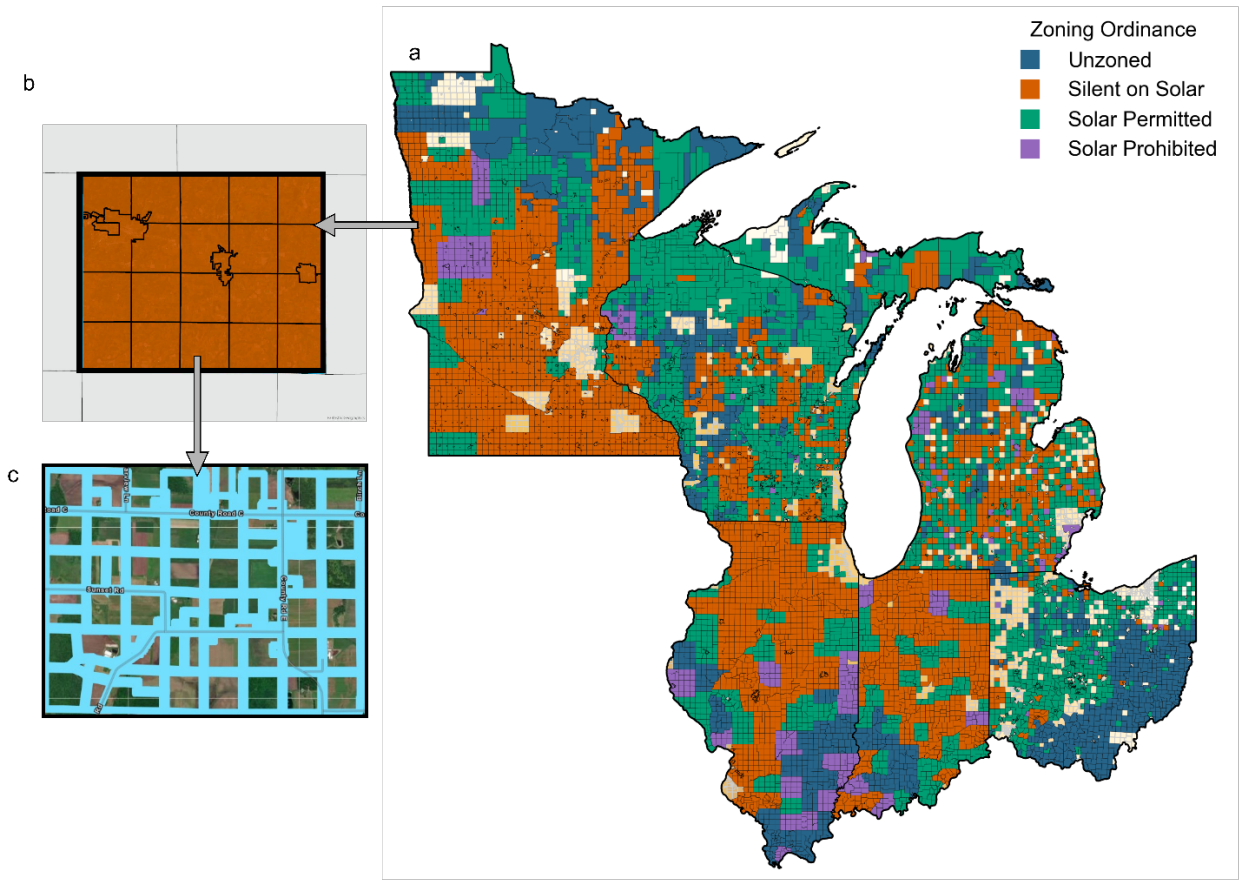

**Fig. S4.** a. Map showing regulations pertaining to utility-scale solar energy systems within subdivisions in the study area, categorized as unzoned, solar permitted in jurisdiction, silent on solar, or solar prohibited in jurisdiction. b. Zoomed-in map showing a county-level zoning ordinance applied across county subdivisions. c. Zoomed-in map showing the implementation of zoning ordinances on land parcels within subdivisions where solar is allowed.

**Table S1.** Input Dataset and Model Assumptions

| Dataset | Source | Input assumptions | URL |
|---|---|---|---|
| Urban Areas | US Census Bureau | Urban areas are excluded from potential siting for utility-scale solar and wind projects. | https://www.census.gov/programs-surveys/geography/guidance/geo-areas/urban-rural.html |
| Urban Buffer(km) | Generated using GIS based on Urban Areas dataset | Buffer distances applied around urban areas:<br>- **Solar**: 0.5 km<br>- **Wind**: 1 km | N/A (Generated dataset) |
| Counties | US Census Bureau | | https://www.census.gov/geographies/mapping-files/time-series/geo/cartographic-boundary.html |
| County subdivisions | US Census Bureau | | https://www.census.gov/geographies/mapping-files/time-series/geo/cartographic-boundary.html |
| Wisconsin Land Parcel Boundaries | Statewide Parcel Map Initiative, Wisconsin State Cartographer's Office | | https://www.sco.wisc.edu/parcels/data/ |
| Indiana Land Parcels Boundaries | Indiana Geographic Information Office (IGIO) | | https://www.indianamap.org/ |
| Slope | 10m Digital Elevation Model | Solar: 10°>, Wind: 19°><br><br>Thresholds Based On: Leslie et al. and Wu et al. | https://earthexplorer.usgs.gov/ |
| Agricultural/cultivated land | 2021 Cropland Data Layer, USDA NASS | Used the cultivated land layer. Excluded non cultivated lands. | https://www.nass.usda.gov/Research_and_Science/Cropland/Release/index.php |
| Prime Farmland | USDA Natural Resources Conservation Service (NRCS) | **Inclusion Criteria**: Lands classified as Prime Farmland, Farmland of Statewide Importance, or Unique Farmland | https://www.nrcs.usda.gov/resources/data-and-reports/web-soil-survey |
| Capacity factor | NREL NSRDB, SAM | **Spatial resolution**: 9km x 9km<br>**Geometry**: point<br>**Design**: Single-axis tracking<br>**System azimuth**: 180°<br>**System losses**: 14%<br>**Inverter efficiency**: 96%<br>**DC to AC**: 1.1 | https://nsrdb.nrel.gov/data-viewer<br><br>https://sam.nrel.gov/ |
| Levelized cost of electricity (LCOE) | NREL System Advisor Model (SAM), 2021 NREL ATB | **Project Lifetime**: 30 years<br>**Discount Rate (Real):** 5.4% (reflecting a WACC for utility-scale projects)<br>**Capital Costs:** From NREL ATB 2021, Moderate Scenario, Low Resource class<br>**O&M Costs**: From NREL ATB 2021, Moderate Scenario<br>**Capacity Factor**: Derived from NSRDB and SAM simulations | https://sam.nrel.gov/<br><br>https://atb.nrel.gov/ |

| | | Degradation Rate: 0.5% per year Tax Assumptions: Investment Tax Credit (30% ITC) Inflation Rate: 2.5% | |
|---|---|---|---|
| Electric transmission lines and substations | US DHS Homeland Infrastructure Foundation-Level Data (HIFLD) | Used to identify proximity to existing transmission infrastructure for interconnection considerations. | https://hifld-geoplatform.opendata.arcgis.com/datasets/electric-power-transmission-lines/data |
| Existing Solar Facilities | USGS National Solar Arrays | | https://eerscmap.usgs.gov/uswtdb/ |
| Generation technology capital and O&M cost | 2021 NREL ATB | Scenario: Moderate Lifetime: 30 years Solar Resource Class: Low Resource (Capacity Factor ≈ 17%–20%) Core Metric Case: Market Technology: Utility-scale PV, Single-axis tracking Capital Expenditures (CAPEX): Includes equipment, installation, and development costs Operational Expenditures (OPEX) Fixed O&M costs from NREL ATB | https://atb.nrel.gov/electricity/2021/ |
| Transmission capital cost | MISO | Cost Estimates: Based on MISO's Transmission Cost Estimation Guide Voltage Levels: Costs vary by transmission line voltage (e.g., 69 kV, 138 kV, 345 kV) Cost Components: Include materials, labor, right-of-way acquisition, and permitting | https://cdn.misoenergy.org/20210209%20PSC%20Item%2006a%20Transmission%20Cost%20Estimation%20Guide%20for%20MTEP21519525.pdf |
| Interregional transmission capacity and load regions | NREL ReEDS | | https://www.nrel.gov/analysis/reeds/ |
| Electricity demand projections | NREL Electrification Futures Study | Scenario: Moderate TMY: 2012 | https://www.nrel.gov/analysis/electrification-futures.html |
| Power densities | Miller and Keith (2018), Environmental Research Letters | Solar: 5.4 W/m² Wind: 0.5 W/m² Mean values for wind power and solar power plants observed in 2016 | DOI 10.1088/1748-9326/aae102 |

**Supplementary Information 2**

# Optimizing Utility-Scale Solar Siting for Local Economic Benefits and Regional Decarbonization:

**PV Solar Net Economic Impact Estimation Model**

Papa Yaw Owusu-Obeng [a, *], Steven R. Miller [b], Sarah Banas Mills [c, d], Michael T. Craig [a, e]

[a] School for Environment and Sustainability, University of Michigan, Ann Arbor, MI 48109, USA
[b] Department of Agricultural, Food, and Resource Economics, Michigan State University, Lansing, USA
[c] Graham Sustainability Institute, University of Michigan, Ann Arbor, MI 48104
[d] A. Alfred Taubman College of Architecture and Urban Planning, University of Michigan, Ann Arbor, MI 48109
[e] Department of Industrial and Operations Engineering, University of Michigan, Ann Arbor, MI 48109, USA

**This PDF file includes:**
Supporting text
Figures S1
Table S1 to S5
SI References

# Contents

## Introduction

When initiating this research, there were varied examples of economic impacts of utility scale PV solar projects with little comparability. This project set out to establish a standard for estimating the economic considerations of utility scale PV projects. The simulation philosophy follows that adopted by the National Renewable Energy Labs (NREL) and their Jobs and Economic Development Impact (JEDI) models of the economics around energy generation projects. The JEDI models have become the standard for measuring economic effect of energy installation projects. However, when this project initiated, NREL did not have a PV JEDI model for PV installation. The simulation model developed here went further than the JEDI models to net out the economic value of lost economic activity on the acres of PV panels that would have been placed in the next most likely use of the acres to be transitioned to PV solar. In other regards, this estimation model was developed with simplifying assumptions over those in comparable JEDI models with the intent to improve non-expert use of the model.

This model is not designed to provide the final definitive estimate of the economic contribution of an intended project, but rather establish expected baselines. It is designed for non-expert use in generating expected net economic benefits of citing commercial PV systems of various sizes. Alternatively, a comprehensive site-specific estimate of the expected economic impact of a PV solar project requires the skillful analysis of a qualified analyst using specifications specific to the planned PV solar development, property proposed under that development and the local economy for which the installation is to be installed and operated. This report documents the use and methods of development of the SETO Economic Simulation Model.

## Economic Simulation Model

There were two goals in developing the economic simulation model. The first was to make a representative, turnkey spreadsheet simulation model that will find wide use by policy makers. The second was to provide an economic simulation platform for geographically profiling socio-optimal citing of PV solar projects. Accordingly, the development of the SETO Economic Simulation Model was guided by a few overriding objectives:

- To encourage wide use, the model should minimize the number of parameters required of the user;
- To facilitate more advanced users, the model should be flexible in assumptions to allow the user to modify the simulation;
- The model should be all encompassing, not requiring the user to populate the spreadsheet with expensive third-party software or data.

# Methods

The economic simulation of utility scale PV solar must account for at least two phases of the project: 1) the installation phase and 2) the ongoing operational phase. A third phase is also reasonable, and that is the teardown phase to be completed at the end of the expected life. Each phase exhibits different expenditure and employment profiles and has different considerations around the timing and sequence of economic activities. In addition, as such projects have a 25-to-30-year life, the models must account for social time value of money preferences and inflation. Social time value of money preferences recognizes the human nature to value benefits today more than the same benefits in the future. Similarly, time value preferences also reveal that we prefer to put off costs rather than experience them now. Economists discount the value of distant benefits and costs according to well established theory of the *time value of money*, as applied in this analysis.

An unrelated concept, which has the potential to impact future measures of value is the presence of price inflation. The U.S. tends to experience persistent and positive inflation over time. Positive inflation indicates that dollars command less spending power over time. Over most of the prior 20 years, inflation was not a significant consideration in economic planning, but the higher rates of inflation since 2019 have put prices progression in the forefront of economic policy. Despite the uncertainties associated with inflation, the modeling approach employed is to abstract from price questions by setting all current and future prices in real dollar terms.

# Project Specification

The user must set some minimal specifications for the project. The minimum specifications include the state and county of the installation and the name plate capacity of the planned project. These minimal specifications will determine the projected soil productivity, the tax treatment and parameters, and the economic multipliers to apply to project expenditures along all phases of the project. The nameplate capacity determines the number of acres required under current technology, where the model power generation per acre of solar panels follows a single axle tracking mounts.

Other parameters have default values that do not depend on the size of the installation. These parameter values can be overridden by the user. For instance, the inverter load ratio is the current technology's market capacity for converting between direct current (DC) power and alternating current (AC). PV solar cells generate DC power, which requires conversion to AC to be connected to the utility grid. The default ratio is set to 1.25 which is a conservative baseline based on current technologies (Anderson, *et. al*, 2022). Similarly, the expected capacity factor and annual downtime are measures of efficiency in converting sunlight to DC power. Capacity factor may be influenced by the expected annual solar radiation at the installation site while the downtime is the expected percent of possible operating time expected to be offline. The default capacity factor is set to 20.1%, indicating that 20.1% of the nameplate capacity is expected to be reached annually (U.S. Energy Information Administration, 2019). The default annual downtime is set to 10% of the potential operating time.[1] The user can also set the projected life of the solar panel installation to 25 years or 30 years, where 30 years is the default. The life of the project may alter the timing of depreciation and net flows of earnings and tax revenues.

In determining the number acres required for the project, the model asserts 10 acres are required per $MW_{ac}$. Scale economies assert this is likely larger for smaller installations and the acres per MW of capacity will decrease with large projects. Total acres include right of way and infrastructure in excess of the required installed PV solar panels. By assumption, about 60 percent of the committed project acres will be under solar panels. Displaced agricultural acres make up all of the acres under solar panels plus an extra two acres per $MW_{ac}$ for access and mandated offsets.

## Installation Phase

The installation phase is a limited duration phase of the project comprising of land preparation, construction, connection and initial deployment. While the required time has some dependence on the size of the PV installation, the developer has a great deal of control over how many months to allocate to the site development. The model defaults on the number of months for installation to 18 months (Bertsch, 2022), though the user can easily adjust this to meet their expectations. Generally, the number of months required for installation has less impact on the estimated economic impact than the size of the installation and the total expected installation expense. However, an elongated installation phase may prolong the lost agricultural output, as the agriculture disruption is calculated on a 12-month year basis. The number of growing seasons disrupted from production is by assumption rounded up to the next year. For example, an installation phase that requires up to 12 months to complete would disrupt a single season of agricultural production, but 13 to 24 months of installation will disrupt two years of agricultural production by assumption. The total number of agricultural acres displaced by the project is determined by the user but defaults to 2/3rds of the project's committed acres.

The model will gather all relevant PV installation expenditures and model how those expenditures are first captured by the local economy and second, how those expenditures spur secondary transactions in the local economy. Expenditure considerations are limited to on-site development expenditures. This omits from consideration soft site development expenditures for lobbying local and state government for development permits and other siting costs incurred in identifying target sites for development. It also excludes offsite engineering and legal expenses.

Initial onsite expenditures by category (Table 1) are estimated as a proportion of total installation cost per $MW_{dc}$ nameplate capacity, currently set at $1.262 per $W_{dc}$ or $1,262 per $MW_{dc}$. The share of costs by component is provided by NREL studies of PV solar development projects (Ramasamy, *et. al*, 2021), with minimal modification. The value of $1.262 per $W_{dc}$ of nameplate capacity is measured in 2022 dollars and can be changed by the user and all the distributed category expenditures will adjust proportionately. Additionally, the user can alter the category distribution of cost shares by changing the share values, but the share values must sum to 100 for the model estimates to be valid.

---

[1] Potential operating time is the number of hours in a year, or approximately 8,768 hours.

| Installation Expenditures by Category | | |
|---|---|---|
| **Installation costs per W$_{dc}$** | **$/W$_{dc}$** | **Share of Total** |
| Engineering, Procurement and Construction Margins | $0.057 | 4.5% |
| Contingency | $0.035 | 2.8% |
| Developer Overhead | $0.028 | 2.2% |
| Transmission Line | $0.014 | 1.1% |
| Interconnection Fee | $0.035 | 2.8% |
| Permitting Fee | $0.014 | 1.1% |
| Sales Tax | $0.057 | 4.5% |
| Engineering, Procurement and Construction Overhead | $0.071 | 5.6% |
| Install Labor & Equipment | $0.156 | 12.4% |
| Electrical Balance of System | $0.099 | 7.9% |
| Structural Balance of System | $0.170 | 13.5% |
| Inverter | $0.057 | 4.5% |
| PV Modules | $0.468 | 37.1% |
| **Total** | **$1.262** | **100%** |

**Table S1: Installation Expenditures by Category**

While on-sight expenditures measure the relevant costs of installing the PV project, not all of those expenditures will be expected to benefit the local economy. Table 2 shows heuristic capture rates, in percentage of total expenditures that are captured by the local economy. The expenditure categories are mapped into those categories in Table 2 and the user can modify the local share of expenditures captured to meet expectations. The share values are bound to lie between and included zero and one. However, this set of parameters may be a primary source of critique by reviewers as there are no restrictions on values the percentage takes. Two notable zero values deserve attention. The contingency cost projection capture is set to zero, not necessarily because the expectation is that such expenditures will go to outside vendors, but more so because allocations to unanticipated expenditures should not be part of an a priori economic impact assessment. By setting the local capture share to zero, the model assures the projections of unanticipated costs are not part of the economic impact assessment. The second zero is PV Modules. Under very rare conditions should this not be zero. Such may exist if the projected solar installation is in a county where PV modules are manufactured.

In the example below, installation is set to 100 percent local research, as installation mostly entails construction activities by which skills are generally readily available. All of these categories can be adjusted, depending on the developer's intentions. The PV developer will often indicate the extent to which they plan to hire services and workers from the local economy but may be limited by availability. As the developer will often contract with local firms for excavation and construction services, as well as for heavy equipment leasing, they may only have indirect control of the local capture despite their intentions.

The parameters provided in Table 2 are based on three sources of information. First, the representative county data was used to determine the extent to which commodities that make up the category are provided in the county. Each of the categories in Table 2 can be made up of multiple commodities and the more of those commodities/services exist, the more assured local availability. The second source was a general heuristic of the generality of the skills required under each category. Construction is largely a general skill that can be undertaken by a relatively larger share of the population than engineering. Accordingly, finding local sources for engineering may be more difficult than that of construction. Finally, local capture estimates from previous economic simulation projects for Michigan PV solar developments and applied to other studies were combined. Collectively, these inputs went into the development of the local share coefficients in Table 2. Of course, the developer may intend to purchase from local suppliers and not find sufficient sources to meet their needs. They can also plan not to employ local suppliers. Table 2 largely reflects local availability, and the developer's stated intention should be taken into account when specifying the final local share capture.

| Local Capture of Install Expenditures by Expenditure Category | Percent |
|---|---|
| Engineering, Procurement and Construction | 25.0% |
| Contingency | 0.0% |
| Transmission/Interconnection | 65.0% |
| State Taxes and Fees | 80.0% |
| Installation (non-residential construction) | 100.0% |
| Electrical components (less transmission) | 15.0% |
| Structures | 25.0% |
| PV Modules | 0.0% |

**Table S2: Local Capture of Installation Expenditures**

The economic impact simulation model will compile the expenditures by category captured in the local economy and estimate the economic effects measured in annual employment equivalence, earnings and wages, contributions to gross regional product and value of sales/output. Note that only employment requires a modifier in interpreting the outcome. This is because the model will estimate the total number of jobs that would accomplish that level of expenditure if it were completed in 12 months' time. Given the potential for the construction phase to span more or less than 12 months, the analysis must recognize that estimated installation job counts represent the number of jobs supported by the expenditure, regardless of how long it takes to spend down that level of expenditure. For example, if it requires one construction worker to complete a $100,000 construction project, then the project should require one such worker whether that was spent in six months, 12 months or 24 months. If completed in six months, it is likely two ½-year workers completed the task, making up one annual equivalent job. If over 24 months, it is likely that one worker working half time over two years would complete the task – once again accounting for single year equivalence job. To be further nuanced, the actual job counts may be less informative than the actual earnings for the installation phase. This is because the employment may be spread over many contractors and workers engaged in temporary work at different times of the year, rather than a limited set of workers, working consistently on this project over the course of a year. That is, one year-equivalent job may be spread out over more than one worker, working part-time or temporarily employed on this project.

More detail on the methods of estimating economic impact can be found below in the Economic Simulation Model section of this report.

## Operations and Maintenance Phase

Operations and maintenance of the project is an ongoing concern with recuring expenditures year-over-year. As is the case with most electrical and mechanical equipment, reliability decreases over time. The NREL provides an estimate of the typical standard operational expenditures of PV solar projects, but this schedule is subject to change over time (Walker, et al., 2020).  Because operational costs are variable over time, the strategy for modeling the annual costs was to establish a benchmark year as the average annual costs over the life of the project and the apply a formula-driven schedule of total annual O&M costs over the 25- or 30-year life of the project that averages to the benchmark year. To minimize the required calculations in accomplishing this, a linear approximation to the NREL-derived schedule (Walker, et al., 2020) was adopted that asserts a five percent annual increase in operating costs at the baseline values at year 15 of the life of the project. Accordingly, each year before year 15, the annual operating costs are expected to be about five percent lower than the next year, while each subsequent year beyond year 15, it is expected to increase by about five percent. Since the growth rate is linearized, the actual year over year change in percentage terms will approximate five percent annual growth.

While the NREL report that provides installation cost estimates provides some baseline O&M cost estimates (Ramasamy, et al., 2021), the O&M cost breakouts are less precise than desired for modeling economic effects. Accordingly, base year O&M costs start with NREL estimates for a single axle tracking system of project size of 100 $MW_{ac}$ or higher, as reported in an alternative NREL report (NREL, 2018). These estimates are dated. Hence baseline estimates were adjusted for technological advances (NREL, 2022, SunSpec Alliance, 2016).

The resulting annual cost profile establishes the baseline aggregate operating costs per $MW_{DC}$ of output per year and breakouts of costs by category. The cost breakouts include payment to lease holders, taxes and other operating costs which may be related to other project characteristics, like the number of acres leased versus purchased, the planned lease rate, the planned inverter ratio and others. Such may have material impacts on the expenditure components and the aggregate costs per MW of generated electricity. Additionally, under other objectives in this USDoE grant, a survey established how leaseholders have used the PV lease payments, including if they save, reinvest, or retire and move away from the community. All of these have implications on how dollars are captured and circulate in the local economy and affect the expected local economic impact of the project.

Table 3 shows the baseline assumptions of the O&M costs and are specified relative to operating output, indicating that tax (in this particular example for Iowa) makes up the largest single cost component to O&M. The land lease costs are the second largest expense, under this example followed distantly by technical inspections and technicians. As not all expenditures are expected to be captured by contractors and businesses in the local economies, expenditures must be assigned a value of local capture. The model has underlying assumptions of the local capture that can be over-ridden by the user.

| Annual Operating Expenditures by Category | $/kW$_{DC}$/year | Share of Total |
|---|---|---|
| **Total O&M Costs per kW$_{DC}$/year** | **$21.690** | |
| **Annual Operating Expenditures by Category** | | |
| Administrator | $2.254 | 10.4% |
| Cleaner | $2.015 | 9.3% |
| Inverter specialist | $0.004 | 0.0% |
| Inspector | $2.443 | 11.3% |
| Journeyman electrician | $1.016 | 4.7% |
| PV module/array Specialist | $2.585 | 11.9% |
| Network/IT | $0.001 | 0.0% |
| Master electrician | $0.475 | 2.2% |
| Mechanic | $0.239 | 1.1% |
| Pest control | $0.044 | 0.2% |
| Structural engineer | $0.000 | 0.0% |
| Mower/Trimmer | $1.623 | 7.5% |
| Utilities locator | $0.005 | 0.0% |
| Land Lease | $4.406 | 20.3% |
| Taxes | $4.580 | 21.1% |
| **Total** | **$21.690** | **100.0%** |

**Table S3: Components of O&M Costs**

Table 4 uses heuristics to assign the share of expenditures captured in the local economy. As pest control and mowing is largely available in even the smallest of counties, the model conjectures that the local economy captures 100 percent of these expenditures. Alternatively, the specialized technical services or centralized operational functions are expected to be provided remotely, such that the local economy will capture none of these expenditures. Other expenditures fall in between, as shown in Table 4.

By assumption in the model, the breakout and local shares do not change over time in the simulation model and there is currently no option for changing these parameters at different stages of the project's life. However, as mentioned above, the total O&M costs will vary from less spent per year to more per year, such that the average annual O&M expenditures will match that shown on the Parameter spreadsheet of the spreadsheet model (Table 3).

| Local Capture of O&M Expenditures by Expenditure Category | Percent |
|---|---|
| Administrator | 0.0% |
| Cleaner | 75.0% |
| Inverter specialist | 25.0% |
| Inspector | 0.0% |
| Journeyman electrician | 50.0% |
| PV module/array Specialist | 20.0% |
| Network/IT | 0.0% |
| Master electrician | 25.0% |
| Mechanic | 25.0% |
| Pest control | 100.0% |
| Structural engineer | 0.0% |
| Mower/Trimmer | 100.0% |
| Utilities locator | 0.0% |
| land lease | 75.0% |
| Taxes | 80.0% |

**Table S4: Local Capture of O&M Expenditures**

The economic simulation model will take these inputs, along with the project specifications to derive a typical annual economic impact estimate of the O&M phase. It will also generate a time series of annual estimated economic impacts to create a schedule of expected annual economic effects, where all values are measured in constant dollars for the year of the analysis. This implies that inflation is not factored into the estimates, making for easier interpretation.[2]

## Teardown Phase

The final economic impact component for estimation is the cost of deconstructing, or decommissioning, the project. Decommissioning generally anticipates restoring the property to pre-project conditions. If in agriculture, that would mean removing all above and below ground structures and restoring drainage tiles and soils to agricultural-use conditions. Most townships and municipalities go so far as to require a performance bond be issued for the decommissioning of the project. Such a bond exists beyond the life of the development or developer to assure funds will be available to restore the property, even if the venture goes bankrupt.

The expected life of these projects is between 25 and 30 years, suggesting that much can happen with energy technology during this time. It is conceivable that technological advances may supplant PV solar, rendering the project obsolete before the planned end of life. However, the most likely outcome is that the project will persist through the planned end of life for the particular project. At that point, the developer may have two options. One is to remove the installation and restore the land to its original condition. The second is to repurpose the installation with a next generation solar energy project – by which we can hardly conceive of how that may look. In this model, we assume the property will be restored to its original condition at the end of the project's life.

Because PV solar projects are relatively new and because such projects have significant lifespans, few examples of the costs of decommissioning exist. However, four recent PV installation proposals included the expected costs of decommissioning the site (Dupuis, 2021; Stantec Consulting Services, 2021; Sandifer, 2021; Terracon, 2021). Many of these studies refer to a 2020 New York study on the required tasks for decommissioning PV sites (NYSERDA, 2020). The estimated costs of decommissioning activities are derived from average decommissioning expenditures per- $MW_{dc}$ nameplate capacity at $41,969 per $MW_{dc}$. The NYSERDA (2020) study provides a breakout of expenditures by category. Collectively these provide a scalable expenditure profile for estimating the economic impacts of decommissioning activities.

Decommissioning generally requires less time than installation. The default number of months required to decommission a site is set to nine months. While the user can change this value, the length of time necessary to decommission a site only impacts our economic impact interpretation if the total number of months exceed a year. Whether the decommissioning requires one month or 12, we assume one growing season is disrupted. Hence, economic net effects of no agricultural output on affected agricultural acres applies.

---

[2] For an example of the implication, $10 an hour for a grounds service worker in 2023 may be equivalent to $15 per hour for the same worker in 2030. In this analysis, the price of workers remains constant at the 2023 prices of 10$ per worker and the audience of the model does not have to conjecture what the value of the dollar in the future will be, as it is set fixed at 2023 prices.

| Decommissioning Expenditure Category | Percent |
|---|---|
| Remove Rack Wiring | 4.1% |
| Remove Panels | 4.1% |
| Dismantle Racks | 20.5% |
| Remove Electrical Equipment | 3.1% |
| Breakup and Remove Concrete Pads or Ballasts | 2.5% |
| Remove Racks | 13.0% |
| Remove Cable | 10.8% |
| Remove Ground Screws and Power Poles | 23.0% |
| Remove Fence | 8.2% |
| Grading | 6.6% |
| Seed Disturbed Areas | 0.4% |
| Truck to Recycling Center | 3.7% |
| **Total Decommissioning** | **100.0%** |

**Table S5: Decommissioning Cost Breakouts**

## Economic Simulation Model

The standard model for measuring macroeconomic-level effects is a representative input-output model for the region in question. There are a few options for creating the representative input-output model. Commercial providers like IMPLAN (2024) and Lightcast (2024), can provide county-specific input-output models. However, county-specific models for the seven-state area encompassing this model, these commercial options would be cost prohibited. Rather, this project selects counties in each of the seven states to be representative of all counties of the same size. Hence, all small counties in the state will then be represented by the randomly selected county chosen in that state.

Input output models trace economic linkages through transactions across the economy. The concept underlying the modern rendering of the input-output model dates to the 1940s and has remained a key tool for understanding macroeconomic impacts for over 60 years. They start with a nationally-specified social accounting matrix (SAM) that is derived from surveys of businesses at the national level. The surveys detail what commodities are purchased by industries and what commodities institutions like households purchase over the course of a year. The SAM entries by industry are double-entry ledgers of transactions that recognize that one industry's purchases are other industries' sales revenue. Both purchases and sales can be made with parties outside the US in the form of imports and exports. Each industry has different regional linkages from which inter-industry transactions arise. That is, each industry exhibits unique expenditure patterns with other industries, institutions and the rest of the World and these expenditure patters depend on availability of inputs by other providers in the local economy.

Input-output models are useful for estimating how a potential change in production or economic activity ripples throughout the economy. Multiplier effects are industry- and geographically specific. Industries with deep local supply chains tend to generate larger macroeconomic effects than industries that primarily rely on suppliers from outside the local economy. Similarly, geographies with larger economies, like metropolitan areas, tend to have larger multiplier effects than economically smaller regions, like rural economies. This is because larger economies tend to capture more transactions internally rather than rely on imports. Coughlin and Mandelbaum (1991) provide a gentle introduction to the workings of input-output models, while Richardson (1985) provides a more rigorous presentation.

The impact simulation starts by properly specifying the direct expenditures, or sources of economic impact. Direct effects are all the transactions that can be directly attributed to the installation, operation and decommissioning of the solar project, depending on the phase. They include new transactions that would not have taken place in the local economy without the project, as well as transactions that do not take place because of the project. For example, construction of the solar panel array on agricultural land necessitates direct expenditures to install the panels but also results in lost agricultural activities on that land. Properly specifying these direct effects is imperative for estimating economic impacts accurately and objectively.

This model devises representative economies to represent all economies of size for a given state. Three size categories were created. Small counties are those with population counts of 15,000 or less. Medium counties have more than small counties, up to 30,000 residents. Finally, large counties are designated as having a population no greater than 50,000 residents but larger than that of medium counties. The selected counties are shown in Figure 1, where eligible counties exclude those currently established by the Office of the Management and Budget as being members of a metropolitan statistical area (U.S. Census. 2022).

The results of the input-output model are reported in terms of Output value, Earnings, Employment generated, and total Value Added. Output value represents the total value of production (all Project activities) as measured as the grand total of receipts generated over the course of a year. Employment and earnings represent the number of jobs involved in the Project and the total labor income shared among them all. Jobs in IMPLAN are calculated as total annual job counts. This means that 1 "job" as estimate by the model may in fact be realized across 2 or more individuals in partial-year roles. This is often the case for highly intensive activities over a short-term project like those anticipated at the installation phase. Earnings are combined proprietors' and wage earnings and includes fringe benefits but excludes taxes and payments to social security. Finally, Value Added is a term that represents total regional income and comprises all labor and proprietors' income, corporate profits and net earnings by government – generally in the form of tax revenues. All positive economic effects measured represent total additional, or "new" economic activities generated by the project that would not otherwise exist. All negative effects represent reductions in economic activities.

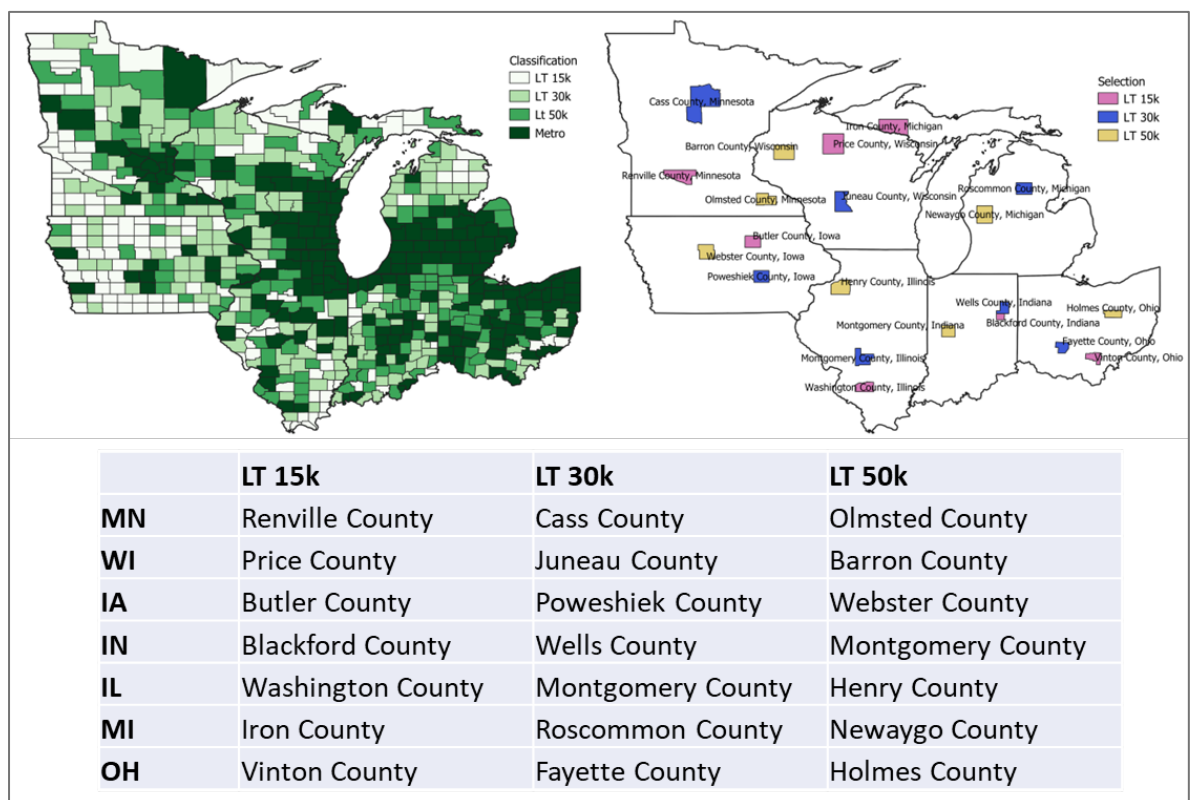

**Figure S1: Representative Geographies**

| | LT 15k | LT 30k | LT 50k |
|---|---|---|---|
| **MN** | Renville County | Cass County | Olmsted County |
| **WI** | Price County | Juneau County | Barron County |
| **IA** | Butler County | Poweshiek County | Webster County |
| **IN** | Blackford County | Wells County | Montgomery County |
| **IL** | Washington County | Montgomery County | Henry County |
| **MI** | Iron County | Roscommon County | Newaygo County |
| **OH** | Vinton County | Fayette County | Holmes County |

## Agricultural Offsets

Economic impact assessments account for all changes in economic activity. The installation, O&M and decommissioning phases of the project have the potential to disrupt existing agricultural production on crop acres converted to solar. In assessing the lost agricultural production, the model assumes the targeted number of acres in agriculture disrupted by the PV installation would have been used to grow a standard three-year rotation of corn (conservative till), soybeans (no till) and wheat (conservative till: grain and straw). This posits a typical row crop operation in the Midwest and acts as a representative agricultural placeholder.

Modeling the economic contribution of crop production starts with the number of agricultural acres impacted by the solar project. Crop enterprise budgets then provide a basis for modeling crop transactions and associated economic impacts of crop production activities through the multiplier effect described above.

Crop enterprise budgets are projected statements of expected per-acre revenues and expenditures and are created for specific crops and rotations. Because the assumed rotation is a three-crop in three-year rotation, we will use a simple average over the three crops in deriving the typical annual crop revenues. We use Ohio State University Extension (2021) enterprise crops for the three-crop rotation. These budgets assume $7.00/bu. corn, $7.50/bu. wheat, and $14.25/bu. Soybeans with yields of 147, 57 and 74 bushels to the acre, respectively. Because

the level of purchases of inputs, not the value of crops sold, is the primary driver of economic impacts, changing the agricultural commodity value will not have a material change on outcomes and therefore, is not a built-in option with the model.

While the model assumes that disrupted agricultural acres are otherwise used for common grain production, converting acres that are in higher-value uses could impose a more significant economic loss to the community than that estimated by the model. Particularly acres in specialty crops like asparagus, peppers, cucumbers, tomatoes or other similar specialty crops require substantially more input and labor during the growing season and command higher market prices. Some agricultural commodity production operations are not good candidates to be converted to PV solar production. For example, converting vineyards and orchards would have similar effects but would also disband years of establishment costs required before the vineyard or orchard is productive. Accordingly, it is unlikely to see such acres converted to solar.

There may be some potential to overstate and to understate the expected loss in agricultural production. There exists a potential to overstate the true expected economic losses of reduced agricultural output for several reasons. First, farms with marginal soils will be more willing to lease their property for non-farm uses. That is, there is a self-selection bias that favors placing low productivity acres into PV use. This may be offset by the developer's pursuit of acres that give direct access to utility trunk lines. Second, it is probable that not all the acres converted to PV will offset agricultural production. For the same reason above, unproductive, or fallow land will likely be converted before productive land is. However, because we assume that all the impacted acres would have been in grain production, it is possible that the lost economic value of foregone agricultural production understates the true economic value if those acres would otherwise be used for high-value specialty crops (tomatoes, strawberries, peppers, etc.), or if the acres are part of livestock operations. The analysis undertaken here may therefore overstate OR understate the value of lost agricultural production. We therefore make assumptions based on what we consider is the likely use of the land in Jackson County, Michigan.

# Optimizing Utility-Scale Solar Siting for Local Economic Benefits and Regional Decarbonization:

Transmission System Modeling

Papa Yaw Owusu-Obeng [a, *], Steven R. Miller [b], Sarah Banas Mills [c, d], Michael T. Craig [a, e]

[a] School for Environment and Sustainability, University of Michigan, Ann Arbor, MI 48109, USA
[b] Department of Agricultural, Food, and Resource Economics, Michigan State University, Lansing, USA
[c] Graham Sustainability Institute, University of Michigan, Ann Arbor, MI 48104
[d] A. Alfred Taubman College of Architecture and Urban Planning, University of Michigan, Ann Arbor, MI 48109
[e] Department of Industrial and Operations Engineering, University of Michigan, Ann Arbor, MI 48109, USA

**This PDF file includes:**

      Supporting text
      Table S1 to S5
      SI References

## 1. Regional transmission modeling

All transmission lines connecting pairs of subregions are aggregated into a single bidirectional line, whose capacity equals the sum of the capacities of the original lines [8]. To simplify calculations, the length of each aggregated line is approximated by the distance between the centroids of the connected subregions. Our capacity expansion (CE) model can increase transmission line capacity to ease congestion and accommodate new generation sources. This expansion is optimized by minimizing the cost per megawatt (MW) of additional transmission capacity, computed as the product of the per-MW-mile cost and the line's length. Within each subregion, the CE model also optimizes the spatially-explicit costs of connecting new facilities to the bulk transmission grid.

## 2. Sub-regional Solar Interconnection cost modelling

To estimate spatially-explicit costs of connecting utility-scale PV facilities to the bulk transmission grid, we refer to the Midcontinent Independent System Operator's (MISO's) transmission cost estimates [14] and follow the ArcGIS workflow outlined in outlined in [15]. We consider utility-scale PV facilities at each subdivision, connecting them by the shortest straight-line distance to either one of five voltage classes (69 kV, 161 kV, 230 kV, 345 kV, and 500 kV). Existing transmission lines are obtained from the Homeland Infrastructure Foundation Level Database [16].

We analyze the cost of two major components of transmission infrastructure: electrical components (conductor and substation costs) and land costs (right-of-way and permitting cost). We estimate the cost of conductors by multiplying the transmission line length by the cost per mile (Table S4.1 in SI). To enable voltage transformation, a new substation is built at the point-of-interconnection. Substation costs are based on their MVA rating (Table S4.4 and S4.5 in SI). We calculate the right-of-way cost by multiplying the total acreage per terrain along the transmission line path by the land preparation cost per acre for that terrain type (Table S4.6 in SI). Finally, we derive the annualized capital cost ($/MW-Year) assuming a project life of 30 years and an interest rate of 3%.

### a. Estimating cost of structures

To estimate the cost of the supporting structures for conductors, we assume that steel towers are used and apply the costs outlined in [4]. Specifically, we calculate the material and installation cost for tangent, angle, and dead-end unit types. Table S4.2 and S4.3 present the total procurement and installation cost for each unit type, along with the number of units required per mile of transmission line, based on the voltage class.

**Table S1**. Structural cost per voltage classes for single circuit transmission lines [15]

| Voltage Class | 69 | 161 | 230 | 345 | 500 |
|---|---|---|---|---|---|
| Tangent (#/mile) | 9 | 7 | 5 | 4.5 | 3 |
| Running Angle (#/mile) | 1 | 1 | 1 | 1 | 1 |
| Non-angled Dead End (#/mile) | 0.25 | 0.25 | 0.25 | 0.25 | 0.25 |
| Angled Dead End (#/mile) | 0.25 | 0.25 | 0.25 | 0.25 | 0.25 |

| Cost per Mile | $528,430 | $520,772 | $580,308 | $969,786 | $1,104,267 |

**Table S2**. Structural cost per voltage classes for double circuit transmission lines [15]

| Voltage Class | 69 | 161 | 230 | 345 | 500 |
|---|---|---|---|---|---|
| Tangent (#/mile) | 9.5 | 7.5 | 7 | 6 | 5 |
| Running Angle (#/mile) | 1 | 1 | 1 | 1 | 1 |
| Non-angled Dead End (#/mile) | 0.25 | 0.25 | 0.25 | 0.25 | 0.25 |
| Angled Dead End (#/mile) | 0.25 | 0.25 | 0.25 | 0.25 | 0.25 |
| Cost per Mile | $849,838 | $1,005,009 | $1,150,818 | $1,991,936 | $2,254,661 |

### b. Estimating cost of building a substation cost

To account for voltage transformation, a new substation is built when connecting generating resource point to an existing transmission line. Our analysis employs MISO's cost estimates for power transformers [14] which includes the unit cost per MVA for each voltage and ampacity rating outlined in Table S4.4. The per unit cost outlined in Table S4.4 includes all material, procurement, and installation costs. To account for operating and maintenance expenses we include an annual cost of $1,543.65 per MVA based on an average cost from 12 utilities [15]. Additionally, our analysis accounts for different substation layouts. As outline in Table S4.5, for a single circuit line, we assume 4 position-breaker-and-a-half bus and for a double circuit line we assume a 6 position-breaker-and-a-half bus for added reliability. We do not include the expenses associated with network-upgrades, such as the cost of upgrading transmission lines and substations, or the expenses related to interconnection studies.

**Table S3**. Power transformer cost ($/MVA)

| Voltage Class | 69 | 161 | 230 | 345 | 500 |
|---|---|---|---|---|---|
| 69 | $4,961 | $4,705 | $5,217 | $6,406 | $8,262 |
| 161 | $4,705 | $6,745 | $5,494 | $6,406 | $7,862 |
| 230 | $5,217 | $5,494 | $7,472 | $6,406 | $7,862 |
| 345 | $6,406 | $6,406 | $6,406 | $9,102 | $8,262 |
| 500 | $8,262 | $7,862 | $7,862 | $8,262 | $12,198 |

**Table S4**. Cost for various substation layouts.

| Voltage Class | 69 | 161 | 230 | 345 | 500 |
|---|---|---|---|---|---|
| 4 positions (breaker-and-a-half bus) | $7.9M | $10.6M | $12.1M | $17.5M | $25.4M |
| 6 positions (breaker-and-a-half bus) | $8.4M | $11.2M | $12.8M | $18.7M | $27.3M |

### c. Estimating right-of-way cost

We calculate the length of each transmission line along each terrain type Table S4.6. Using the length per terrain type and the right-of-way width, we obtain the total acreage per terrain type. We multiple this value by the land preparation cost per terrain to obtain the right of way cost. Additionally, we include an overhead of $15,235 per acre to account for permitting and land acquisition.

**Table S5**. Classification of land cover types and corresponding land preparation cost

| Terrain Type | NLCD Classes (Land Cover) | Preparation Cost ($/Acre) |
|---|---|---|
| light vegetation | 21, 22, 23, 24, 31, 51, 52, 71, 72, 73, 74, 81, 82 | $272 |
| forest | 41, 42, 43 | $5,577 |
| wetland | 11, 90, 95 | $111,865 |
| mountain | 12 | $7,169 |

### d. Total Implementation Costs

To estimate the total project cost, we analyze various overhead cost items, such as Administrative and General Overhead (5.5%), Project Management Cost (1.5%), and Allowance for Funds Used During Construction (AFUDC) estimated at 7.5%. To accommodate uncertainty in the cost per mile estimation method, we apply a 30% contingency [14]. To calculate the annualized capital cost ($/MW-Year), we derive the capital recovery factor (CRF) at 5.1%, assuming a conservative project life of 30 years and an interest rate of 3%. Furthermore, we include an annualized operating and maintenance benchmark of $1,543.65 per MVA, based on the average cost of substations from 12 utilities [15]. For transmission lines, an annualized operating cost of $7,300.75 per circuit mile, based on the average from 17 utilities. Both benchmarks are adjusted for inflation from 2009 to 2020 using the consumer price index.

**SI References**

**Supplementary Information 5**

# Optimizing Utility-Scale Solar Siting for Local Economic Benefits and Regional Decarbonization:

## Results

Papa Yaw Owusu-Obeng [a, *], Steven R. Miller [b], Sarah Banas Mills [c, d], Michael T. Craig [a, e]

[a] School for Environment and Sustainability, University of Michigan, Ann Arbor, MI 48109, USA
[b] Department of Agricultural, Food, and Resource Economics, Michigan State University, Lansing, USA
[c] Graham Sustainability Institute, University of Michigan, Ann Arbor, MI 48104
[d] A. Alfred Taubman College of Architecture and Urban Planning, University of Michigan, Ann Arbor, MI 48109
[e] Department of Industrial and Operations Engineering, University of Michigan, Ann Arbor, MI 48109, USA

**This PDF file includes:**

Results

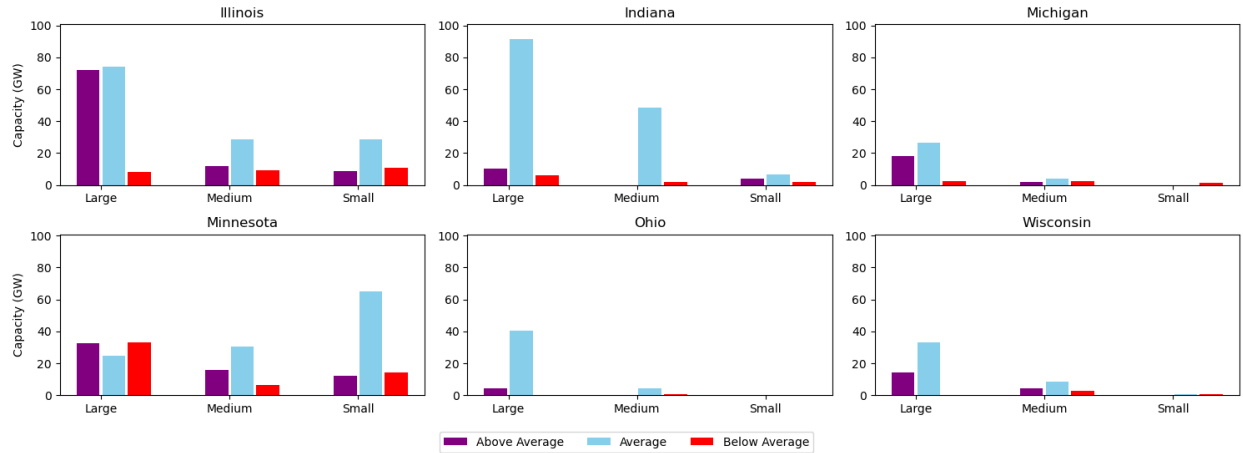

Figure S1. Distribution of total utility-scale solar potential (GW) by economy size (Large, Medium, Small) and agricultural productivity level of converted land (Above Average, Average, Below Average) across states.

Table S1. Average net property taxes per state across counties

| State | Property Tax in Year 1 | Net present value |
|-------|------------------------|-------------------|
| Ohio | 8,750 | 103,400 |
| Wisconsin | 4,980 | 60,300 |
| Indiana | 6,080 | 39,400 |
| Michigan | 4,750 | 38,800 |
| Illinois | 3,690 | 28,000 |
| Minnesota | 1,670 | 19,700 |

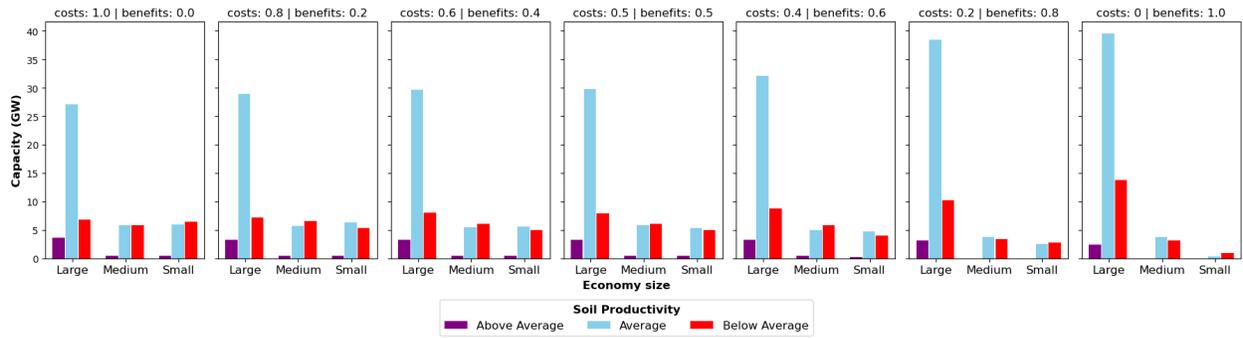

Figure S2. State-level solar investments by economy size and soil productivity under different cost-benefit scenarios

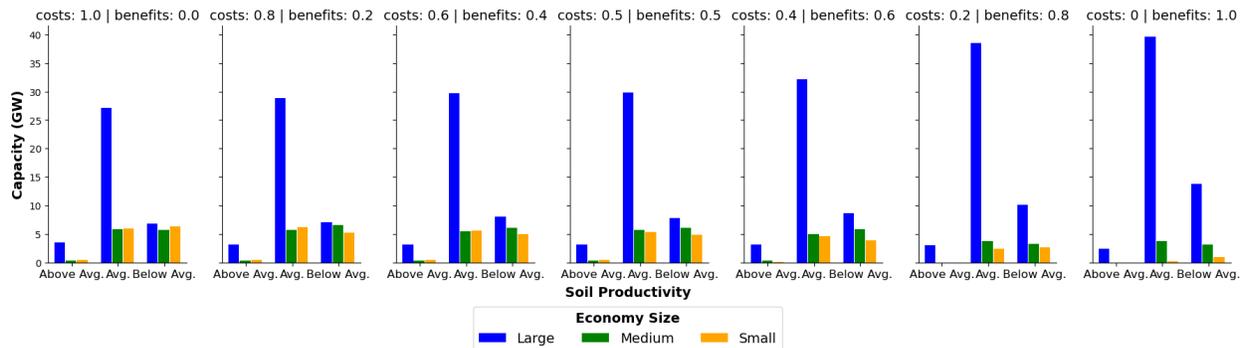

Figure S3. State-level solar investments by soil productivity and economy size under different cost-benefit scenarios

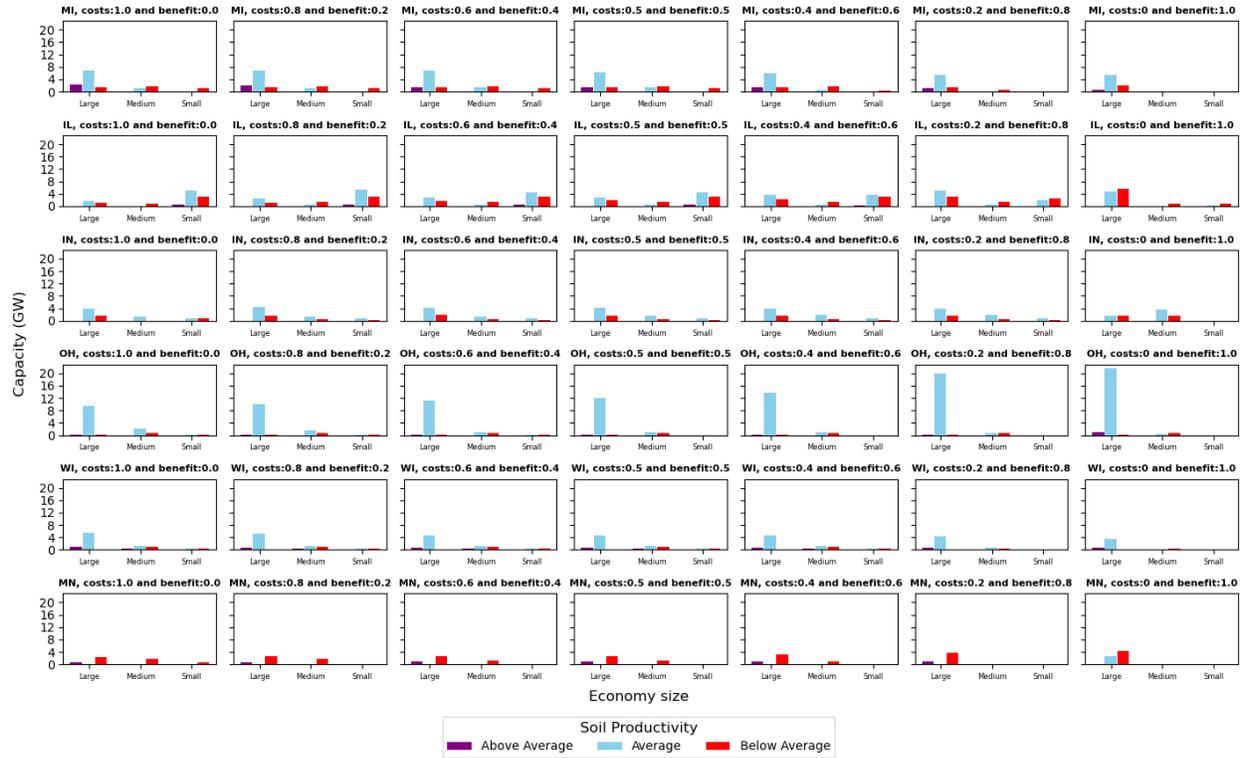

Figure S4. Solar investments at the state-level by economy size and soil productivity for different cost-benefit scenarios

Table S2. Summary of the total fixed and variable solar investments developed in the study region and the contribution to value-added from the share of investments captured locally.

| | Solar Economic Impact (Billion USD) | Total investment cost | Change in benefit: 100% cost vs 80% benefit (Million USD) |
|---|---|---|---|
| Scenario | | | State |
| 100% : 0% | 8.70 | $378.00 | Minnesota: 442 (41%) |
| 80% : 20% | 8.80 | $378.20 | Ohio: 520 (24%) |
| 60% : 40% | 9.00 | $378.30 | Indiana: 75 (5%) |
| 50% : 50% | 9.10 | $378.30 | Illinois: 44 (3%) |
| 40%: 60% | 9.20 | $378.40 | Wisconsin: 16 (1%) |
| 20% : 80% | 9.70 | $379.70 | Michigan: -100 (-7%) |
| 0% : 100% | 10.10 | $381.80 | |

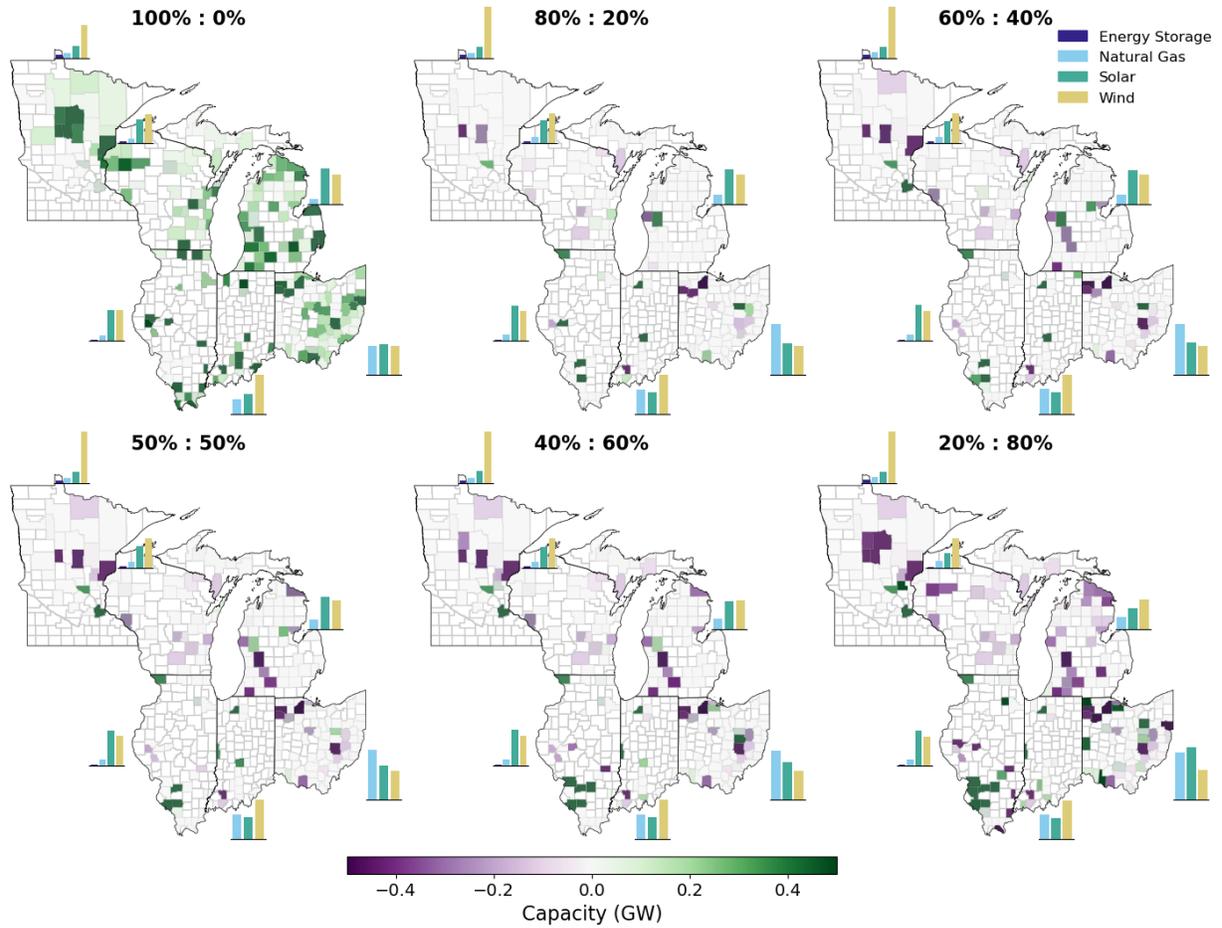

Figure S5. Geographic shifts in new utility-scale solar as the weighting ratio between cost to local economic benefit is adjustment (local economic benefit increases from top left to bottom right). Positive values on the scale (green) represent capacity additions, while negative values (purple) indicate capacity reductions in states. The colorbars indicate capacity differences for all technologies, with green for solar, yellow for wind, purple for energy storage, and blue for natural gas.